\documentclass[prb,aps,tightenlines,twocolumn,
showpacs,floatfix]{revtex4}
\usepackage{graphicx}
\usepackage{amssymb}
\usepackage{amsmath}
\usepackage{bm}
\usepackage{colordvi}
\usepackage{mathrsfs}

\begin{document}
\title{Optical response of graphene under
  intense terahertz fields} 
\author{Y. Zhou}
\author{M. W. Wu}
\thanks{Author to whom correspondence should be addressed}
\email{mwwu@ustc.edu.cn.}
\affiliation{Hefei National Laboratory for Physical Sciences at
  Microscale and Department of Physics, University of Science and
  Technology of China, Hefei, Anhui, 230026, China}

\date{\today}
\begin{abstract}
 Optical responses of graphene in the presence of intense circularly
  and linearly polarized terahertz fields are investigated based
  on the Floquet theory. 
  We examine the energy spectrum and density of states. 
  It is found that gaps open in the quasi-energy spectrum due to
  the single-photon/multi-photon resonances.
  These quasi-energy gaps are pronounced at small momentum, but
  decrease dramatically with the increase of momentum and finally tend
  to be closed when the momentum is large enough. 
  Due to the contribution from the states at large momentum, the gaps
  in the density of states are effectively closed, in contrast to
  the prediction in the previous work by Oka and Aoki [Phys. Rev. B 
  {\bf 79}, 081406(R) (2009)]. 
  We also investigate the optical conductivity for different field 
  strengths and Fermi energies, and show
  the main features of the dynamical Franz-Keldysh effect in
  graphene. It is discovered that the optical conductivity exhibits a
  multi-step-like structure due to the sideband-modulated optical
  transition. It is also shown that dips appear at 
  frequencies being the integer numbers  of the
  applied terahertz field frequency in the case of low Fermi energy,
  originating from the quasi-energy gaps at small momentums. 
  Moreover,  under a circularly polarized terahertz field,
  we  predict peaks in the middle of the ``steps''
  and peaks induced by the contribution from the states around
  zero momentum in the optical conductivity. 
\end{abstract}



\maketitle

\section{Introduction}
Since the experimental realization of graphene,\cite{Novoselov_04}
a monolayer of carbon atoms arranged in a honeycomb lattice, this material has
aroused enormous interest due to its unique physical
characteristics.\cite{Geim_rev_nat,Neto_rev_09,Beenakker_rev,Orlita_rev,
Chakraborty_rev,Peres_rev,Mucciolo_rev,Sarma_rev,Ferreira_dc}
Among different works in this field, the linear
optical property of graphene is one of the
main focuses of attention.\cite{Orlita_rev,Gusynin,Ando_02,Neto_06,
Stauber_disorder,Stauber_NN,Giuliani_ee,exp_Basov,exp_Nair,exp_APL}
The theoretical works based on the Dirac Hamiltonian show
that the optical conductivity at high frequency is dominated by the
interband optical conductivity, which takes a constant value of
$\sigma_0=\frac{e^2}{4\hbar}$ for frequency larger than twice of the Fermi
energy $E_F$ and approaches to zero for frequency below $2E_F$ due to the Pauli
blocking.\cite{Gusynin,Ando_02,Neto_06,Stauber_disorder,Giuliani_ee}
The universal value $\sigma_0$ of optical conductivity has been demonstrated
to be valid not only in the noninteracting limit,\cite{Ando_02,Gusynin} but also
in the presence of the disorder and the electron-electron interaction
as long as the Dirac-cone approximation remains 
valid.\cite{Neto_06,Stauber_disorder,Giuliani_ee} 
The constant optical conductivity in a wide range of frequency also has been
observed in the optical experiments.\cite{exp_Basov,exp_Nair,exp_APL}
On the other hand, the optical conductivity at low frequency is
determined by the intraband optical conductivity, which presents a Drude peak
centred at zero frequency and is strongly influenced by the sample-dependent
scattering behavior.\cite{Gusynin,Ando_02,Neto_06,Stauber_disorder} 

Recently, influence of an intense ac field on electrical and optical properties in
graphene has also attracted much attention.
\cite{Gupta_dis2,Naumis_pol,Syzranov_08,Oka_cur,Zhang_energy,Fistul_07,
Kibis_quantiz,Oka_opt,Mikhailov_multi,Wright_09,Oka_cur2,Ryzhii_dc,Ryzhii_dcac,
Ryzhii_BG,Wright_BG,Abergel_BG,Abergel_BG2}
It has been found that the application of an intense ac field can dramatically
modify the band structure and hence the density of states
(DOS).\cite{Oka_cur,Oka_cur2,Kibis_quantiz,Zhang_energy,Naumis_pol,Syzranov_08}
Their results also showed that a stationary energy gap appears around the Dirac
point under a circularly polarized ac field.\cite{Oka_cur,Kibis_quantiz}
Oka and Aoki\cite{Oka_cur,Oka_cur2,Oka_opt} calculated the dc and ac
conductivities in graphene irradiated by an intense circularly polarized light via
the extended Kubo formula, and proposed the photovoltaic Hall effect, which is
a novel Hall effect occuring in the absence of uniform magnetic fields.
The dc transport properties of graphene-based p-n junctions under an intense ac
field were also investigated theoretically.\cite{Fistul_07,Syzranov_08}
However, contribution in the optical spectra in graphene
from the optical sidebands\cite{Kono_sideband,Phillips_sideband,
Maslov_sideband,Nordstrom_exp_EDFK,Holthaus_Stark,Rodriguez_Stark,
Yacoby_68,Jauho_DFK,Chin_exp_20,Zhang_exp_06,Srivastava_exp_DFK}
has not yet been well investigated.

In semiconductors, the contribution from the sidebands has been
demonstrated to be important for the optical and transport
properties. Many interesting phenomena, such as the photon-assisted
tunneling,\cite{Hanggi_98,Hanggi_05} 
the sideband generation of exciton,\cite{
Kono_sideband,Phillips_sideband,Maslov_sideband,Nordstrom_exp_EDFK} the ac Stark
effect,\cite{Holthaus_Stark,Nordstrom_exp_EDFK,Rodriguez_Stark} and
the dynamical Franz-Keldysh (DFK) effect,\cite{Yacoby_68,
Jauho_DFK,Chin_exp_20,Zhang_exp_06,Srivastava_exp_DFK} as well as spin
generation and manipulation utilized by the intense terahertz (THz)
field\cite{Cheng_APL,Jiang_JAP,Zhou_PE,Jiang_PRB_07,Jiang_PRB_08}
are related to the formation of sidebands.
Among these effects, the DFK effect describes the influence on optical
spectra from the sidebands of the expanded states,\cite{Yacoby_68,Jauho_DFK}
which includes finite absorption below the band edge from the contribution of
the sidebands below the bottom of the conduction band and the step-like behavior
above the band edge due to the sideband-modulated generalized DOS.
Just as the DFK effect in semiconductors, the formation of optical sidebands
should also influence the optical spectra near the absorption edge around
$2E_{\rm F}$ in graphene. Nevertheless, the band structure of graphene is
gapless and the energy dispersion is linear, which is quite distinct from semiconductors.
Thus the DFK effect in graphene is expected to present some unique
behaviors. This makes the investigation on this problem become very interesting
and important. It is also noted that in previous investigation\cite{Oka_opt} only the
optical conductivity with $E_{\rm F}$ much smaller than the frequency of the
ac field was discussed and thus the contribution from the optical sidebands is
difficult to identify. In the present work, we calculate the optical
conductivity of graphene under the intense THz field for various Fermi energies
and field strengths in order to gain a complete view of the DFK effect
in graphene.

In order to include the contribution from the optical sidebands explicitly,
we solve the time-dependent Schr\"odinger equation by using the
Floquet theory\cite{Shirley_65} and obtain the optical conductivity in
graphene under an intense THz field via the nonequlibrium Green
functions.\cite{Jauho_DFK,Haug_08}
In this paper, we focus on the optical conductivity at high frequency,
which is known to be insensitive to the scattering
strength.\cite{Orlita_rev,Gusynin,Ando_02,Neto_06,
Stauber_disorder,Stauber_NN,Giuliani_ee,exp_Basov,exp_Nair,exp_APL}
This allows us to ignore the detail of the scattering and
only discuss the optical conductivity in the noninteracting limit.
We first examine the energy spectrum and DOS. 
It is found that gaps appear in the quasi-energy
spectrum due to the single-photon/multi-photon
resonances.\cite{Holthaus_Stark,Hanggi_98} These quasi-energy gaps
are pronounced at small momentum, in consistence with the previous
investigations.\cite{Oka_cur,Kibis_quantiz,Syzranov_08} 
However, the quasi-energy gaps decrease dramatically with the increase
of momentum and finally disappear when the momentum is large enough. 
Therefore after taking account of the contribution from the states
with large momentum, gaps in the DOS are effectively closed,
in contrast to the prediction by Oka and Aoki.\cite{Oka_cur}
Our results of the optical conductivity reveal the main features
of the DFK effect in graphene.

This paper is organized as follows. In Sec.~IIA, we obtain the energy spectrum
by exploiting the Floquet theory. Then in Sec.~IIB, we derive the
optical conductivity via the nonequlibrium Green functions. The numerical
results of the DOS and optical conductivity are presented in Sec.~III. Finally,
we summarize in Sec.~IV.

\section{Model and Formalism}
\subsection{Hamiltonian}
\label{Hami}
We consider a graphene layer placed in the $x$-$y$ plane.
In the vicinity of the Dirac point, the effective Hamiltonian of
graphene can be written as ($\hbar\equiv 1$)\cite{DiVincenzo}
\begin{eqnarray}
  \hat{H}_{0}^\mu({\bf k})&=&v_{\rm F}(\mu\hat{\sigma}_x k_x
  + \hat{\sigma}_y k_y).
  \label{H_0}
\end{eqnarray}
Here $\mu=1(-1)$ for $K(K^\prime)$ valley; $v_{\rm F}$ is the Fermi velocity;
${\bf k}$ represents the two-dimensional wave vector relative to $K(K^\prime)$
point; $\hat{\bm{\sigma}}$ is the Pauli matrix in the pseudospin
space formed by the A and B sublattices of the honeycomb lattice.
Here and hereafter, symbols with 
$\hat{\mbox{}}$ present
the $2\times2$ matrices in the pseudospin space.
The eigenvalue and eigenvector of $\hat{H}_0^\mu$ are
$E_{{\bf k}\nu}^\mu=\nu v_{\rm F}|{\bf k}|$ and
$\zeta_{{\bf k}\nu}^\mu=1/\sqrt{2}(\mu\nu e^{-i\mu\theta_{\bf k}},1)^{\rm T}$, respectively,
with $\nu$ being 1 $(-1)$ for electron (hole) band and $\theta_{\bf k}$
representing the polar angle of ${\bf k}$.
Substituting  ${\bf k}$ by ${\bf k}+e{\bf A}(t)$, one obtains
the effective Hamiltonian in the presence of a THz field  %
\begin{eqnarray}
  \hat{H}_{\rm eff}^\mu({\bf k},t)&=&\hat{H}_{0}^\mu({\bf k})
  +\hat{H}_{\rm THz}^\mu(t),\\
  \label{H_eff}
  \hat{H}_{\rm THz}^\mu(t)&=&ev_{\rm F}[\mu\hat{\sigma}_x A_x(t)
  + \hat{\sigma}_y A_y(t)].
  \label{H_THz}
\end{eqnarray}
For convenience, we choose the THz field as
${\bf E}(t)=E_0({\bf e}_x\cos\theta_{\bf E}\cos\Omega t
+{\bf e}_y\sin\theta_{\bf E}\sin\Omega t)$. Thus
the vector potential ${\bf A}(t)$ reads
${\bf A}(t)=\frac{E_0}{\Omega}(-{\bf e}_x\cos\theta_{\bf E}\sin\Omega t
+{\bf e}_y\sin\theta_{\bf E}\cos\Omega t)$. For $\theta_{\bf E}=0$,
$\pi/4$ and $\pi/2$, the THz
fields are linearly polarized along the $x$-axis, circularly polarized and
linearly polarized along the $y$-axis, respectively.
Without loss of generality, we set the THz field linearly polarized along
the $x$-axis when discussing the case for a linearly polarized THz field.

By exploiting the Floquet theory,\cite{Shirley_65} the solution of the
Schr\"odinger equation
$i\partial_t|\Phi_{{\bf k}\alpha}^\mu(t)\rangle=\hat{H}_{\rm eff}^\mu({\bf k},t)
|\Phi_{{\bf k}\alpha}^\mu(t)\rangle$ has the form (
$|\Phi_{{\bf k}\alpha}^\mu(t)\rangle$
is referred to as the Floquet state in the following):
\begin{equation}
  |\Phi_{{\bf k}\alpha}^\mu(t)\rangle=e^{-i\mathcal{E}_{{\bf k}\alpha}^\mu t}
  \sum_{n=-\infty}^\infty e^{in\Omega t}|\psi^{n}_{\mu{\bf k}\alpha}\rangle,
  \label{Phi_def}
\end{equation}
in which $\alpha$ represents the branch index of the
solution; $\mathcal{E}_{{\bf k}\alpha}^\mu$
and $|\psi^{n}_{\mu{\bf k}\alpha}\rangle=(\psi^{n+}_{\mu{\bf k}\alpha},
\psi^{n-}_{\mu{\bf k}\alpha})^{\rm T}$ are the eigenvalue (quasi-energy) and
eigenvector determined by
\begin{eqnarray}
  \nonumber
&&\hspace{-0.8cm} (\frac{\mathcal{E}_{{\bf k}\alpha}^\mu}{\Omega}-n)
\psi^{n\sigma}_{\mu{\bf k}\alpha}=
  \frac{i\beta}{2} [ (\mu \cos\theta_{\bf E}-\sigma\sin\theta_{\bf E})
  \psi^{n-1\,-\sigma}_{\mu{\bf k}\alpha}
  \\ &&\hspace{-0.8cm} \mbox{}
  - (\mu \cos\theta_{\bf E}+\sigma\sin\theta_{\bf E})
  \psi^{n+1\,-\sigma}_{\mu{\bf k}\alpha} ]  
  + \frac{v_{\rm F}k}{\Omega}\mu e^{-i\mu\sigma\theta_{\bf k}}
  \psi^{n\,-\sigma}_{\mu{\bf k}\alpha},
  \label{Eq1_eigen}
\end{eqnarray}
with $\beta={v_{\rm F} eE_0}/{\Omega^2}$.
The above equation shows that the relation between the
normalized quasi-energy ${\mathcal{E}_{{\bf k}\alpha}^\mu}/{\Omega}$
and the normalized momentum ${v_{\rm F}{\bf k}}/{\Omega}$
is only determined by the dimensionless quantity $\beta$.\cite{other_def}

Due to the periodicity of $\hat{H}_{\rm eff}^\mu({\bf k},t)$, the eigenvalues
are also periodic, i.e., if $\mathcal{E}_{{\bf k}\alpha}^\mu$ is a solution of
Eq.~(\ref{Eq1_eigen}), then $\mathcal{E}_{{\bf k}\alpha}^{\mu;l}=
\mathcal{E}_{{\bf k}\alpha}^\mu+l\Omega$ is also a solution.
It is evident that the eigenvectors of $\mathcal{E}_{{\bf k}\alpha}^{\mu;l}$
and $\mathcal{E}_{{\bf k}\alpha}^\mu$ satisfy $\psi^{l;n\sigma}_{\mu{\bf k}\alpha}=
\psi^{n-l\,\sigma}_{\mu{\bf k}\alpha}$, thus 
$\mathcal{E}_{{\bf k}\alpha}^{\mu;l}$ and $\mathcal{E}_{{\bf k}\alpha}^\mu$
correspond to the same physical solution of the Schr\"odinger
equation. Namely, the quasi-energy is a multi-valued
quantity of the Floquet state.\cite{Faisal_dis2}
Nevertheless, for each momentum and each valley, the number 
of the independent quasi-energies is 2, which is determined by the
dimension of the Hilbert space. 
For convenience, we choose the independent quasi-energies
$\varepsilon_{{\bf k}\eta}^\mu$ in the reduced Floquet zone
$(-\Omega/2,\Omega/2]$, with $\eta=\pm$ representing the index of
independent solutions. These quasi-energies are referred to as the
reduced quasi-energies in the following. The corresponding
eigenvectors are labelled as $|\phi^{n}_{\mu{\bf k}\eta}\rangle$. 
Therefore, by choosing the proper integer $l$, arbitrary quasi-energy 
$\mathcal{E}^\mu_{{\bf k}\alpha}$ and the corresponding eigenvectors 
$\psi^{n\sigma}_{\mu{\bf k}\alpha}$ can be written into the form
\begin{eqnarray}
  &&\mathcal{E}^\mu_{{\bf k}\alpha}=\varepsilon^\mu_{{\bf k}\eta} +l\Omega,
  \label{relation_1}
  \\ &&\psi^{n\sigma}_{\mu{\bf k}\alpha}=\phi^{n-l\,\sigma}_{\mu{\bf k}\eta}.
  \label{relation_2}
\end{eqnarray}
In addition, for the reduced quasi-energies with different
  $\eta$, one has 
\begin{eqnarray}
  \varepsilon_{{\bf k}+}^\mu&=&-\varepsilon_{{\bf k}-}^\mu,
  \label{symmetry_1}\\
  |\phi^{n}_{\mu{\bf k}+}\rangle&=&\sigma_y|\phi^{-n}_{\mu{\bf
      k}-}\rangle^\ast.
  \label{symmetry_2}
\end{eqnarray}

From Eq.~(\ref{Phi_def}), it is seen that the Floquet state 
$|\Phi_{{\bf k}\eta}^\mu(t)\rangle$ (the general solution index
$\alpha$ has been replaced by the independent solution index $\eta$)
contains Fourier components with different 
frequencies, quite distinct from the eigenstate of 
$H_0$, which only takes one eigen-frequency.
Each Fourier component corresponds to a sideband of this Floquet state.
For the Floquet states with reduced quasi-energy 
$\varepsilon^\mu_{{\bf k}\eta}$, sidebands 
appear at the frequencies (quasi-energies)
\begin{equation}
{\xi}_{{\bf k}\eta}^{\mu;n}=\varepsilon_{{\bf k}\eta}^\mu-n\Omega 
\label{energy_side}
\end{equation}
and the corresponding weights are
\begin{equation}
  W^{\mu;n}_{{\bf k}\eta}=\langle
  \phi^{n}_{\mu{\bf  k}\eta}|\phi^{n}_{\mu{\bf k}\eta}\rangle.
  \label{weight_side}
\end{equation}
From Eqs.~(\ref{symmetry_1}) and (\ref{symmetry_2}), one can see that 
the quasi-energies and weights of the sidebands satisfy
\begin{eqnarray}
  {\xi}_{{\bf k}+}^{\mu;n}&=&-{\xi}_{{\bf k}-}^{\mu;m}-(n+m)\Omega,\\
  \label{symmetry_3}
  W^{\mu;n}_{{\bf k}+}&=&W^{\mu;-n}_{{\bf k}-}.
  \label{symmetry_4}
\end{eqnarray}

Besides the quasi-energy, another important quantity of the Floquet state is the
mean energy\cite{Gupta_dis2,Faisal_dis2,Hsu_dis2,Martinez_meanenergy} 
\begin{eqnarray}
  \nonumber
  \overline{\varepsilon}_{{\bf k}\eta}^\mu&=&\frac{1}{T_0}\int^{T_0}_0dt\langle
  \Phi_{{\bf k}\eta}^\mu(t)| \hat{H}_{\rm eff}^\mu |\Phi_{{\bf k}\eta}^\mu(t)\rangle\\
  &=&\varepsilon_{{\bf k}\eta}^\mu-\sum_{n\sigma}n\Omega
  {\phi_{\mu{\bf k}\eta}^{n\sigma\ast}} \phi_{\mu{\bf k}\eta}^{n\sigma},
  \label{mean_energy}
\end{eqnarray}
where $T_0=2\pi/\Omega$ is the period of the applied THz field.
Independent of the choice of the quasi-energy, the mean energy is a single-valued
structure quantity of the Floquet state. Thus in previous
works,\cite{Gupta_dis2,Faisal_dis2,Hsu_dis2}
the mean energy was utilized to identify the filled Floquet states, i.e.
states with lower mean energy will be occupied at first.
In this paper, we also use this ansatz to obtain the distribution function
of the Floquet state, which will be discussed in the next subsection.
Moreover, the mean energy is used to identify whether the Floquet state is
electron- or hole-like. Analogous to the definition of the electron and
hole states in the field-free case, we define the quasi-electron
($\eta=+$) and quasi-hole states ($\eta=-$) as the Floquet states satisfying
$\overline{\varepsilon}_{{\bf k}+}^\mu>0$ and
$\overline{\varepsilon}_{{\bf k}-}^\mu<0$, respectively.

\subsection{Optical conductivity}
It is known that the optical absorption is measured by the real part of optical
conductivity.\cite{Orlita_rev,Gusynin,Ando_02,Neto_06,
Stauber_disorder,Stauber_NN,Giuliani_ee,exp_Basov,exp_Nair,exp_APL}
Therefore we focus on the real part of the optical conductivity in the
following.
For the probing light field of frequency $\omega_l$ with the polarization in the
$l$ $(=x,y)$ direction, the linear-response theory yields the real part of optical
conductivity:
\begin{equation}
  {\rm Re}\sigma_{ll}(T,\omega_l)=-\frac{g_s g_v}{\omega_l}
  {\rm Im}\Pi^r_{ll}(T,\omega_l).
  \label{Resigma}
\end{equation}
Here $g_v=2$ and $g_s=2$ are the valley and spin degeneracies, respectively;
$\Pi_{ll}^r(T,\omega_l)$ is retarded current-current correlation function in the $K$
valley. Here and hereafter we only give the correlation
function in the $K$ valley and omit the valley index
$\mu$ in all symbols, as the contributions to optical conductivity
from both valleys are identical. 
$\Pi_{ll}^r(T,\omega_l)$ can be written as
\begin{equation}
  \Pi^r_{ll}(T,\omega_l)=\int_{-\infty}^\infty{\rm d}\tau\, e^{i\omega_l\tau}
  \Pi^r_{ll}(T+\frac{\tau}{2},T-\frac{\tau}{2}),
  \label{Pi_omega1}
\end{equation}
where
\begin{equation}
  \Pi_{ll}^r(t,t') = -{i}\theta (t-t') \langle
  [ j^{l}(t),j^{l}(t')]\rangle
\end{equation}
with $j^{l}=ev_{\rm F}\sigma_l$ presenting the $l$ component of the current
operator in graphene.\cite{Ando_02,Neto_06}
Via the nonequlibrium Green function method,\cite{Jauho_DFK,Haug_08}
we have
\begin{eqnarray}
  \nonumber
  \Pi_{ll}^r(t,t') &=& {-i} \hspace{-0.1cm}
  \sum_{{\bf k}\sigma_1\sigma_2 \atop \sigma_1'\sigma_2'} \hspace{-0.1cm}
  [j^{l}_{{\bf k}\sigma_1\sigma_2} G^r_{{\bf k}\sigma_2\sigma_1'}(t,t')
  j^{l}_{{\bf k}\sigma_1'\sigma_2'} G^<_{{\bf k}\sigma_2'\sigma_1}(t',t)  \\
  &&\hspace{-0.2cm} \mbox{}
  + j^{l}_{{\bf k}\sigma_1\sigma_2} G^<_{{\bf k}\sigma_2\sigma_1'}(t,t')
  j^{l}_{{\bf k}\sigma_1'\sigma_2'} G^a_{{\bf k}\sigma_2'\sigma_1}(t',t)],
  \label{Pi_tt}
\end{eqnarray}
with $G^{r(a,<)}_{{\bf k}\sigma_1\sigma_2}(t,t')$ representing the
retarded (advanced, lesser) 
single-particle Green function.\cite{Haug_08}
Substituting Eq.~(\ref{Pi_tt}) into Eq.~(\ref{Pi_omega1}), one has 
\begin{eqnarray}
 \nonumber
\Pi_{ll}^r(T,\omega_l )
&=& {-i} \sum_{{\bf k}\sigma_1\sigma_2 \atop \sigma_1'\sigma_2'}
\int_{-\infty}^\infty {{\rm d}\omega\over 2\pi}
j^{l}_{{\bf k}\sigma_1\sigma_2} j^{l}_{{\bf k}\sigma_1'\sigma_2'}
G^<_{{\bf k}\sigma_2'\sigma_1}(T,\omega) \\
&&\hspace{-0.85cm} \mbox{}\times
\big[ G^r_{{\bf k}\sigma_2\sigma_1'}(T,\omega+\omega_l)
+G^a_{{\bf k}\sigma_2\sigma_1'}(T,\omega-\omega_l) \big].
\end{eqnarray}
Since $\left[G_{{\bf k}\sigma_1\sigma_2}^<(t,t')\right]^\ast = -
G_{{\bf k}\sigma_2\sigma_1}^<(t',t)$ and
$\left[G_{{\bf k}\sigma_1\sigma_2}^a(t,t')\right]^\ast =
G_{{\bf k}\sigma_2\sigma_1}^r(t',t) $,
$\Pi_{ll}^r(t,t')$ is real. Thus one obtains
\begin{eqnarray}
\nonumber
&&\hspace{-0.cm} 
{\rm Im}\Pi_{ll}^r(T,\omega_l) =
{1\over 2i}[\Pi_{ll}^r(T,\omega_l ) - \Pi_{ll}^r(T,-\omega_l )]\\
\nonumber &&\hspace{-0.cm} = 
 {i\over 2} \sum_{{\bf k}\sigma_1\sigma_2 \atop \sigma_1'\sigma_2'} \hspace{-0.cm}
\int_{-\infty}^\infty {{\rm d}\omega\over 2\pi}
j^{l}_{{\bf k}\sigma_1\sigma_2} j^{l}_{{\bf k}\sigma_1'\sigma_2'}
\big[ A_{{\bf k}\sigma_2\sigma_1'}(T,\omega+\omega_l)
\\
&& 
\hspace{0.35cm} \mbox{}
-A_{{\bf k}\sigma_2\sigma_1'}(T,\omega-\omega_l) \big]
G^<_{{\bf k}\sigma_2'\sigma_1}(T,\omega).
\label{ImPi_w}
\end{eqnarray}
In above equations, we have used the relation 
$A_{{\bf k}\sigma_1\sigma_2}(T,\omega)= i[ G_{{\bf k}\sigma_1\sigma_2}^r(T,\omega)
-G_{{\bf k}\sigma_1\sigma_2}^a(T,\omega) ]$.
In the noninteracting limit, the retarded Green function is given by
\begin{equation}
  \hat{G}^r_{\bf k}(t_1,t_2)=-i\theta(t_1-t_2)\sum_{\eta}|\Phi_{{\bf k}\eta}(t_1)
  \rangle \langle\Phi_{{\bf k}\eta}(t_2)|.
\end{equation}
Thus the spectral function in the frequency space can be written as
\begin{eqnarray}
  \nonumber
  \hat{A}_{{\bf k}}(T,\omega)
  &=&\int_{-\infty}^\infty{\rm d}\tau\, e^{i\omega\tau}
  \sum_{\eta}|\Phi_{{\bf k}\eta}(T+\frac{\tau}{2})\rangle
  \langle\Phi_{{\bf k}\eta}(T-\frac{\tau}{2})|\\
  \nonumber
  &=&  2\pi\sum_{nm \eta} e^{i(n-m)\Omega T} 
  |\phi_{{\bf k}\eta}^{n} \rangle \langle \phi_{{\bf k}\eta}^{m}|
  \delta\big(\omega -\varepsilon_{{\bf k}\eta}+(n \\ &&
  \mbox{} +m)\frac{\Omega}{2}\big).
  \label{A_omega}
\end{eqnarray}

The next step is to calculate $\hat{G}_{\bf k}^<(T,\omega)$.
The equal-time lesser function can be expressed in the form
\begin{equation}
  \hat{G}^<_{{\bf k}}(t,t)= i\sum_{\eta_1\eta_2}
  \rho_{{\bf k}\eta_1\eta_2}(t)
  |\Phi_{{\bf k}\eta_1}(t)\rangle \langle\Phi_{{\bf k}\eta_2}(t)|.
  \label{G_tt}
\end{equation}
Here $\rho_{{\bf k}\eta_1\eta_2}(t)$ is the nonequilibrium
density matrix in the Floquet picture,\cite{Jiang_PRB_08,Kohler_PRE}
which can be determined by the kinetic equation including
the electron-impurity, electron-phonon and electron-electron
scatterings.\cite{Haug_08,Jiang_PRB_08,Kohler_PRE,Wu_rev} 
This approach is very complicated and left as the subject of our future work.
Here we obtain the lesser function based on a simple ansatz following the
previous works:\cite{Gupta_dis2,Faisal_dis2,Hsu_dis2} 
the steady-state density matrix is diagonal with the diagonal term 
$f_{{\bf k}\eta}$ being the Fermi distribution on the mean energy of the 
Floquet state,
\begin{equation}
  \rho_{{\bf k}\eta_1\eta_2}(t)= f_{{\bf k}\eta}\delta_{\eta_1\eta_2}
  =n_{\rm F}(\overline{\varepsilon}_{{\bf k}\eta})\delta_{\eta_1\eta_2}.
  \label{dis_2}
\end{equation}
In the present paper, we focus on the case at zero temperature, and then
\begin{equation}
  f_{{\bf k}\eta}=\theta(\overline{\varepsilon}_{{\bf k}\eta}-E_{\rm F}),
\end{equation}
where the Fermi energy $E_{\rm F}$ can be obtained from
\begin{equation}
  N_e =g_sg_v\sum_{{\bf k}\eta}\theta(E_{\rm F}-
  \overline{\varepsilon}_{{\bf  k}\eta}).
\end{equation}
Substituting Eq.~(\ref{dis_2}) into Eq.~(\ref{G_tt}), one gets
\begin{equation}
  \hat{G}^<_{{\bf k}}(t,t)= i\sum_{\eta} f_{{\bf k}\eta}
  |\Phi_{{\bf k}\eta}(t)\rangle \langle\Phi_{{\bf k}\eta}(t)|.
  \label{G_tt_2}
\end{equation}
By further exploiting the generalized Kadanoff-Baym Ansatz,\cite{Haug_08} the
two-time lesser function can be obtained as
\begin{eqnarray}
  \nonumber
  \hat{G}^<_{{\bf k}}(t,t')&=&i\hat{G}^r_{{\bf k}}(t,t')
  \hat{G}^<_{{\bf k}}(t',t')
  - i\hat{G}^<_{{\bf k}}(t,t)\hat{G}^a_{{\bf k}}(t,t')\\
  &=&i\sum_\eta f_{{\bf k}\eta}
  |\Phi_{{\bf k}\eta}(t)\rangle \langle\Phi_{{\bf k}\eta}(t')|.
\end{eqnarray}
Then we have
\begin{equation}
  \hat{G}^<_{{\bf k}}(T,\omega_l)=i\sum_\eta f_{{\bf k}\eta}
  \hat{A}_{{\bf k}\eta}(T,\omega),
\end{equation}
with $\hat{A}_{{\bf k}\eta}(T,\omega)=\int_{-\infty}^\infty{\rm d}\tau\,
e^{i\omega\tau} |\Phi_{{\bf k}\eta}(T+\frac{\tau}{2})\rangle
\langle\Phi_{{\bf k}\eta}(T-\frac{\tau}{2})|$.
Using the above equations, one obtains the optical conductivity
\begin{eqnarray}
  \nonumber
  {\rm Re}\sigma_{ll}(T,\omega_l)
  &=& \frac{g_sg_v\pi}{\omega_l}\hspace{-0.2cm}
  \sum_{{\bf k}\eta_1\eta_2 n_1 \atop m_1 n_2 m_2}\hspace{-0.1cm}
  \langle\phi_{{\bf k}\eta_2}^{m_2}| \hat{j}^l_{{\bf k}}
  |\phi_{{\bf k}\eta_1}^{n_1}\rangle \langle\phi_{{\bf k}\eta_1}^{m_1}|
  \hat{j}^l_{{\bf k}} |\phi_{{\bf k}\eta_2}^{n_2}\rangle \\
  \nonumber
  && \hspace{-0.6cm} \mbox{}\times (f_{{\bf k}\eta_1}-f_{{\bf k}\eta_2})
  {\rm e}^{i(n_1-m_1+n_2-m_2)\Omega T} \delta(\varepsilon_{{\bf k}\eta_1}\\
  && \hspace{-0.6cm} \mbox{}-\varepsilon_{{\bf k}\eta_2}
  -(n_1+m_1-n_2-m_2)\frac{\Omega}{2}+w_l).
  \label{opt_T}
\end{eqnarray}
The time-averaged optical conductivity can be
written as
\begin{eqnarray}
  \nonumber
  &&\hspace{-0.cm}
  {\rm Re}{\sigma}^{\rm ave}_{ll}(\omega_l)=
  \frac{1}{T_0}\int_0^{T_0}dT \;{\rm Re}\sigma_{ll}(T,\omega_l)\\
  \nonumber
  &&\hspace{-0.cm}  
  =\frac{g_sg_v\pi}{\omega_l}\hspace{-0.cm}
  \sum_{{\bf k}\eta_1\eta_2 \atop N n_1 m_1}\hspace{-0.cm}
  \langle\phi_{{\bf k}\eta_2}^{n_1-N}| \hat{j}^l_{{\bf k}}
  |\phi_{{\bf k}\eta_1}^{n_1}\rangle  \langle\phi_{{\bf k}\eta_1}^{m_1}|
  \hat{j}^l_{{\bf k}} |\phi_{{\bf k}\eta_2}^{m_1-N}\rangle \\
  && \hspace{0.35cm} \mbox{} \times 
  (f_{{\bf k}\eta_1}-f_{{\bf k}\eta_2})
  \delta(\varepsilon_{{\bf k}\eta_1}-\varepsilon_{{\bf k}\eta_2}
  -N{\Omega}+w_l).
  \label{opt_ave}
\end{eqnarray}
It is found that the time-averaged optical conductivity from the nonequlibrium
Green functions agrees with that from the extended Kubo formula used in
Ref.~\onlinecite{Oka_opt} (see also Appendix~\ref{modified_oka}).
However, it is noted that our method not only gives the time-averaged optical
conductivity but also the time-dependent one, and thus can
provide the dynamical information of optical response, which can be observed
via the time-resolved measurements. 
It is also noted the distribution used by Oka and Aoki,\cite{Oka_opt} i.e.,
 $f_{{\bf k}\eta}^{\rm ini}=\sum\limits_{\nu l}n_{\rm F}
(E_{{\bf k}\nu})|\langle\zeta_{{\bf k}\nu}|\phi_{{\bf k}\eta}^{l}\rangle|^2$,
is obtained by projecting the equilibrium Fermi
  distribution in the field-free case to the basis set formed by 
 the Floquet eigenvectors, corresponding to the ``sudden approximation''.
Obviously, this distribution, referred to as the projected
distribution in the following,
is quite different from the mean-energy-determined distribution used
in this paper. The comparison of the optical conductivities obtained
from these two distributions will be addressed in next section. 

\begin{figure}[tbp]
  \begin{center}
    \includegraphics[width=7.2cm]{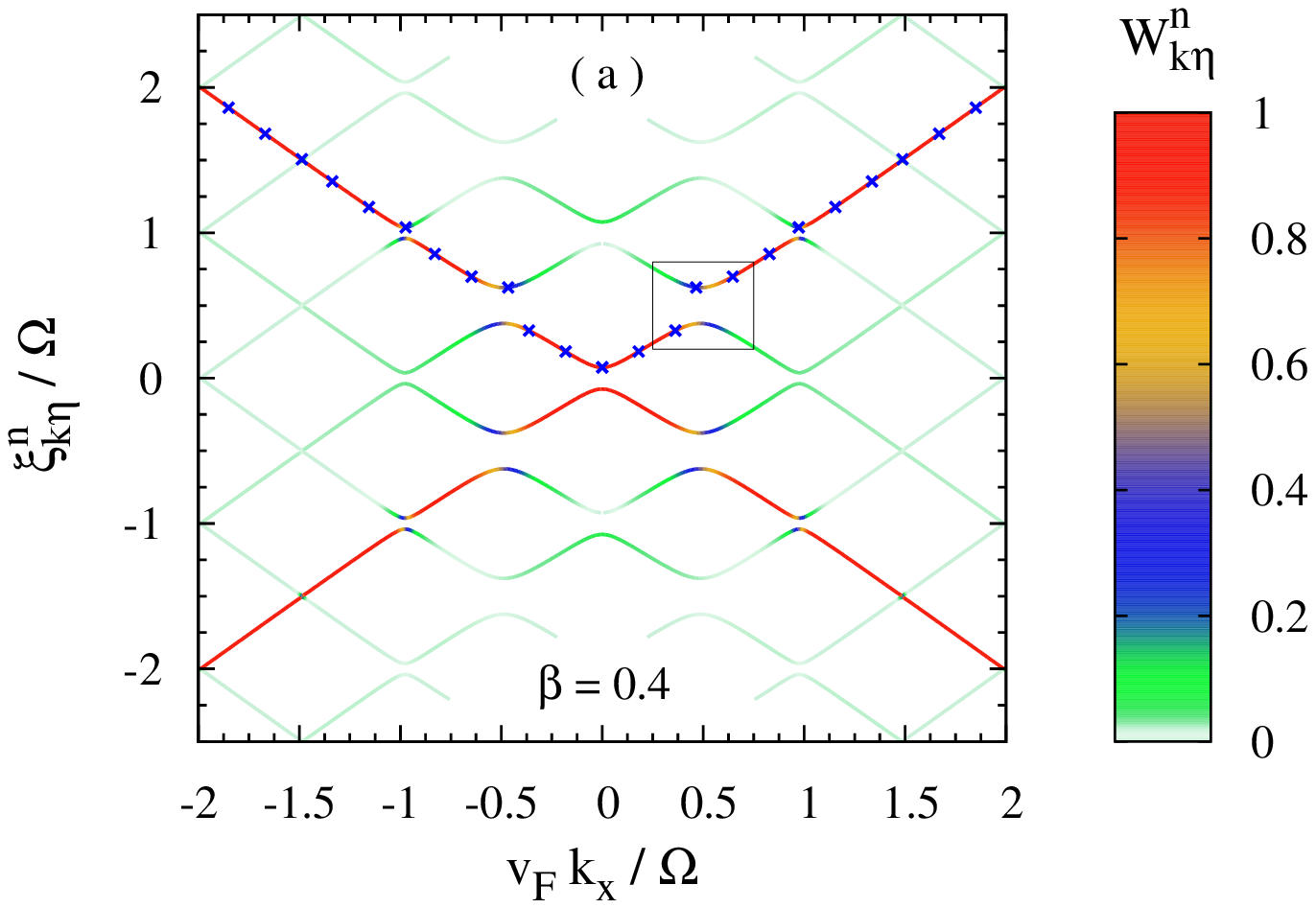}
  \end{center}
  \begin{center}
    \includegraphics[width=6.2cm]{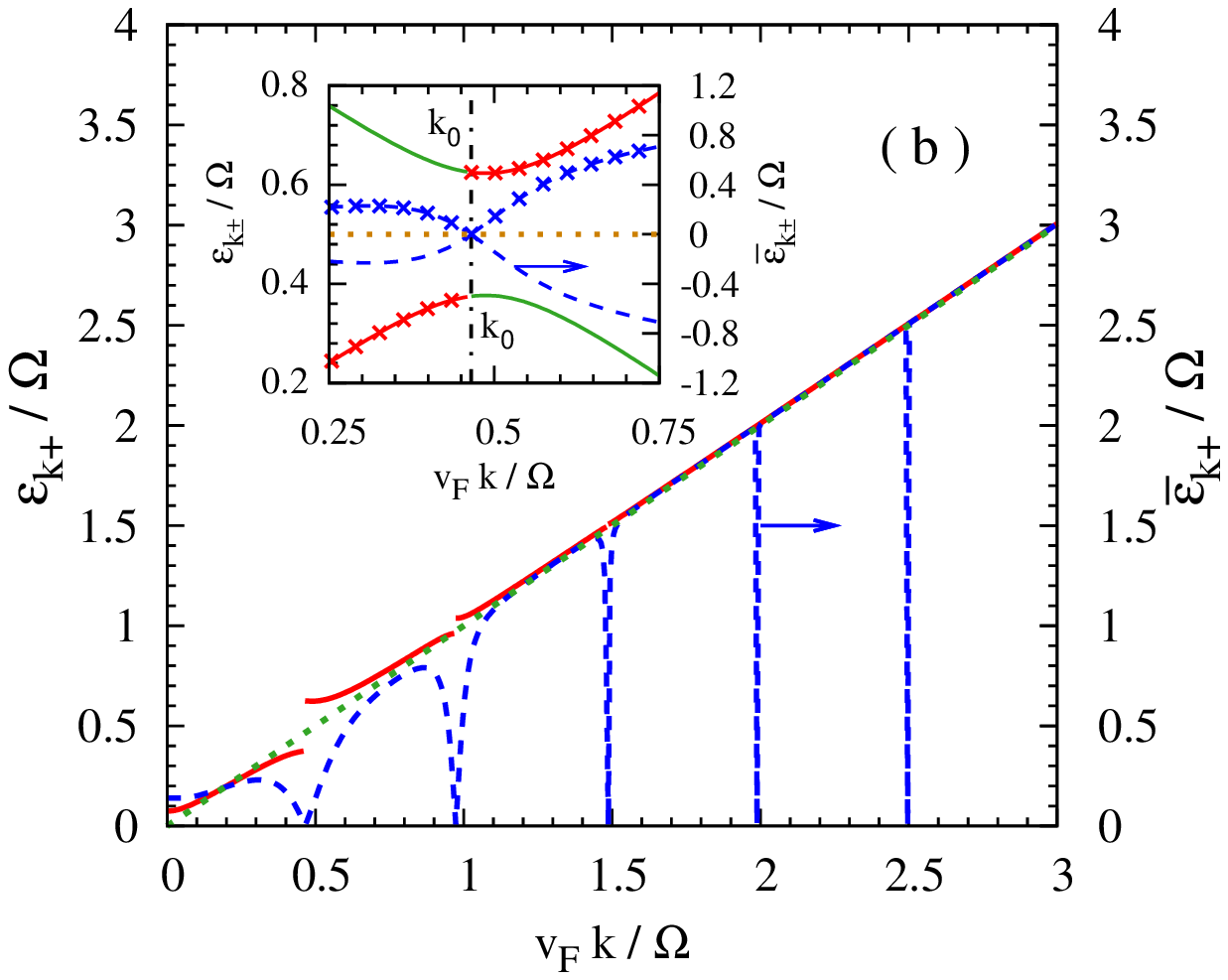}
  \end{center}
  \caption{ (Color online) Circularly polarized THz field
    with $\beta=0.4$. (a) Quasi-energies ${\xi}_{{\bf k}\eta}^{n}$
    against the normalized momentum. The color coding
    represents the weight $W_{{\bf k}\eta}^n$ of the corresponding sideband. 
    The blue crosses represent the selected quasi-energies of the
    quasi-electron states.
    (b) Quasi-energy ${\varepsilon}_{{\bf k}+}$ (red solid curve) and
    mean energy $\overline{\varepsilon}_{{\bf k}+}$ 
    (blue dashed curve) of the quasi-electron state as well as the
    field-free electron energy (green dotted curve)
    against the normalized momentum. 
    In the inset of (b), we plot the quasi-energy around the gap at
    $k=0.5\Omega/v_{\rm F}$ and 
    ${\varepsilon}_{{\bf k}\eta}=0.5\Omega$ [the region labelled by the 
    box in (a)] as well as the corresponding mean energy as function
    of momentum. The solid (dashed) curves with and without crosses 
    represent the quasi-energies (mean energies) of the quasi-electron
    ($\eta=+$) and quasi-hole ($\eta=-$) states, respectively. The
    black chain line indicates the momentum of the crossover point
    $k_0$. Note the scale of mean energy is on the right-hand side of
    the frame.
  }
  \label{fig_energy_weak}
\end{figure}

\begin{figure}[tbp]
  \begin{center}
    \includegraphics[width=7.cm]{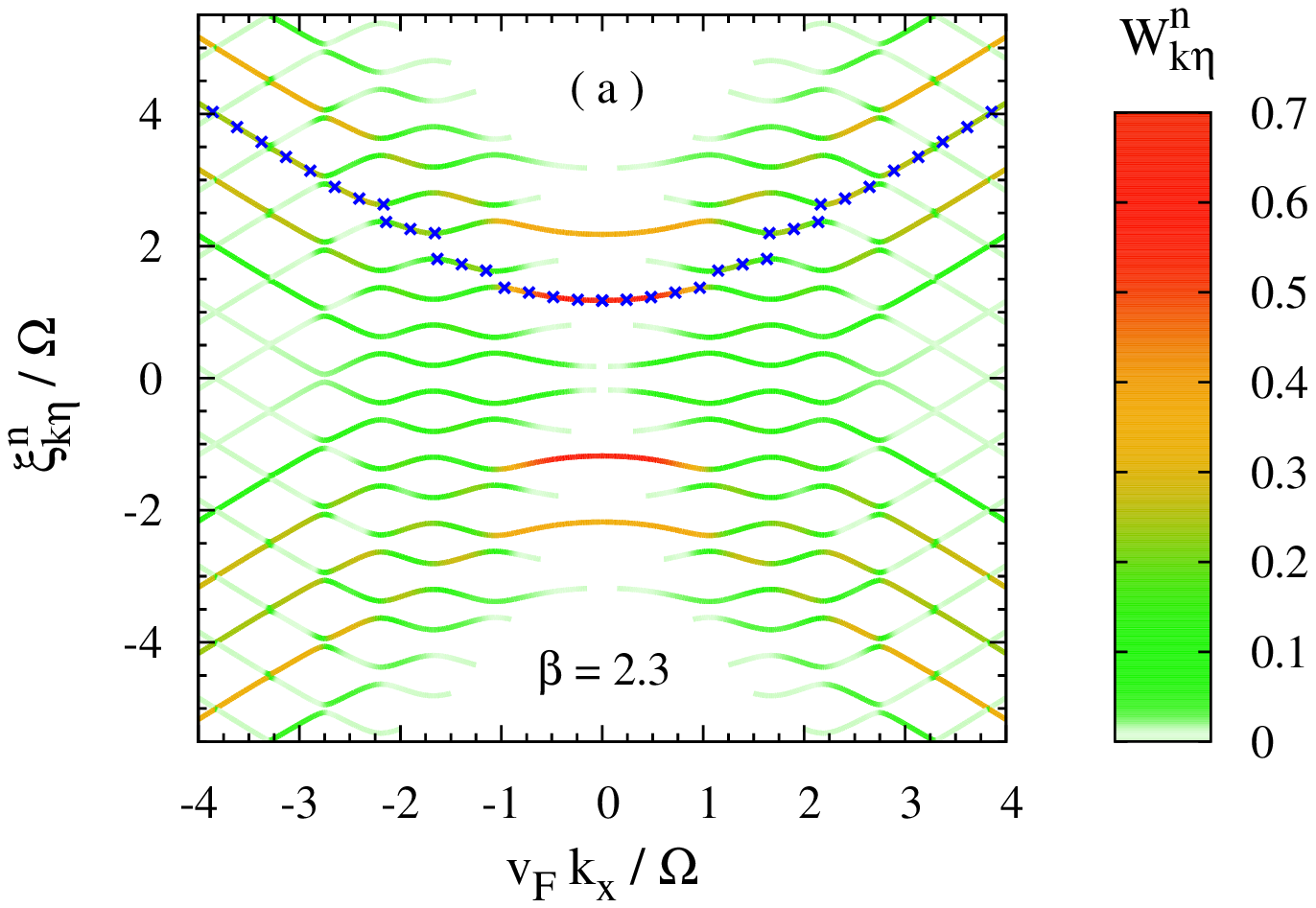}
  \end{center}
  \begin{center}
    \includegraphics[width=6.2cm]{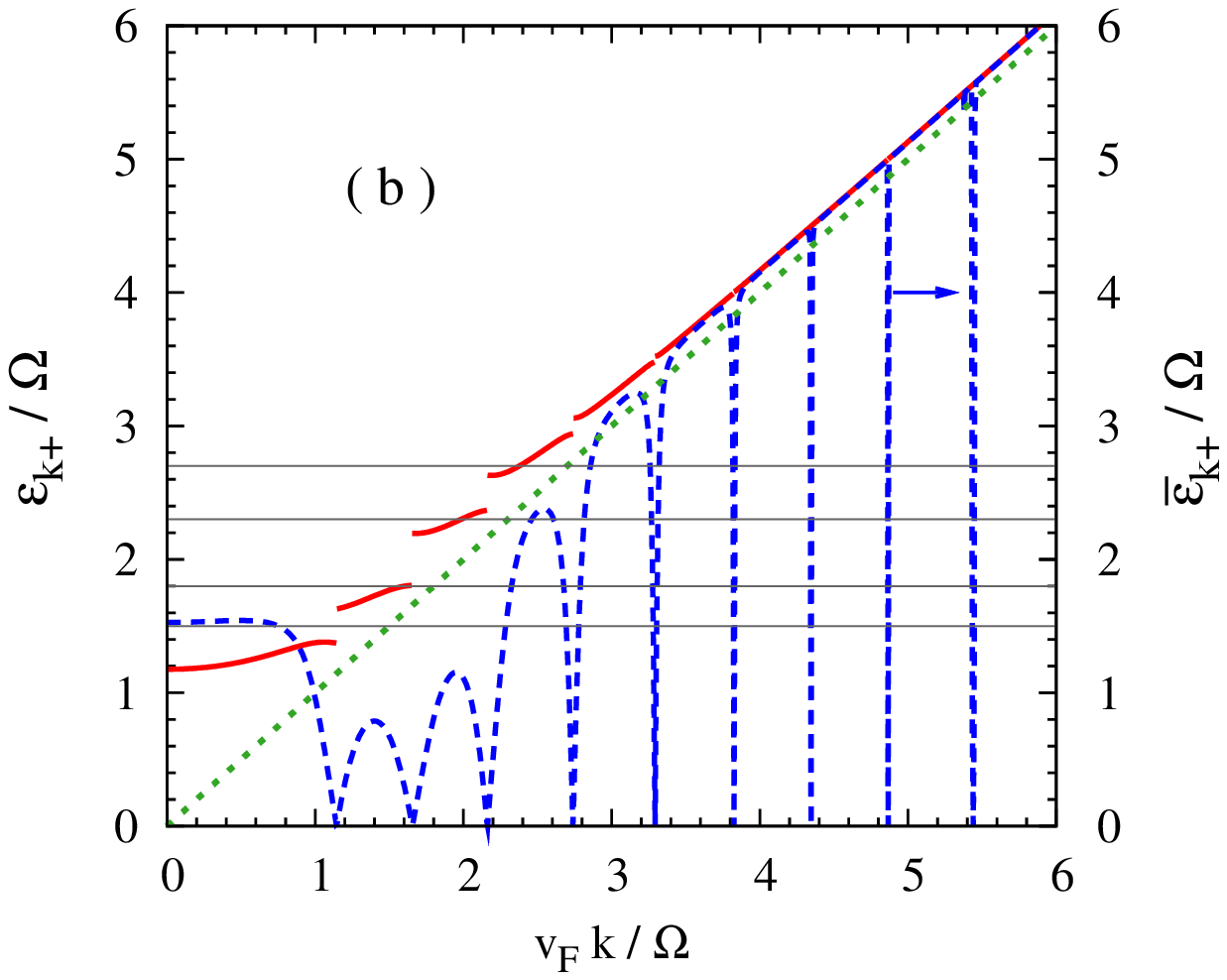}
  \end{center}
  \caption{ (Color online) Circularly polarized THz field
    with $\beta=2.3$. (a) Quasi-energies ${\xi}_{{\bf k}\eta}^{n}$
    against the normalized momentum. The color coding represents
    the weight $W_{{\bf k}\eta}^n$ of the corresponding sideband.  
    The blue crosses represent the selected quasi-energies of the
    quasi-electron states.
    (b) Quasi-energy ${\varepsilon}_{{\bf k}+}$ (red solid curve) and
    mean energy $\overline{\varepsilon}_{{\bf k}+}$ 
    (blue dashed curve) of the quasi-electron state as well as the
    field-free electron energy (green dotted curve) 
    versus the normalized momentum.
    The thin black lines mark the mean energies being $1.5\Omega$,
    $1.8\Omega$, $2.3\Omega$ and $2.7\Omega$, which are used in the 
    Fig.~\ref{fig_cir_lEf}(c). 
    Note the scale of mean energy is on the right-hand side 
    of the frame.
  }
  \label{fig_energy_strong}
\end{figure}

\begin{figure*}[tbp]
  \begin{center}
    \includegraphics[width=17cm]{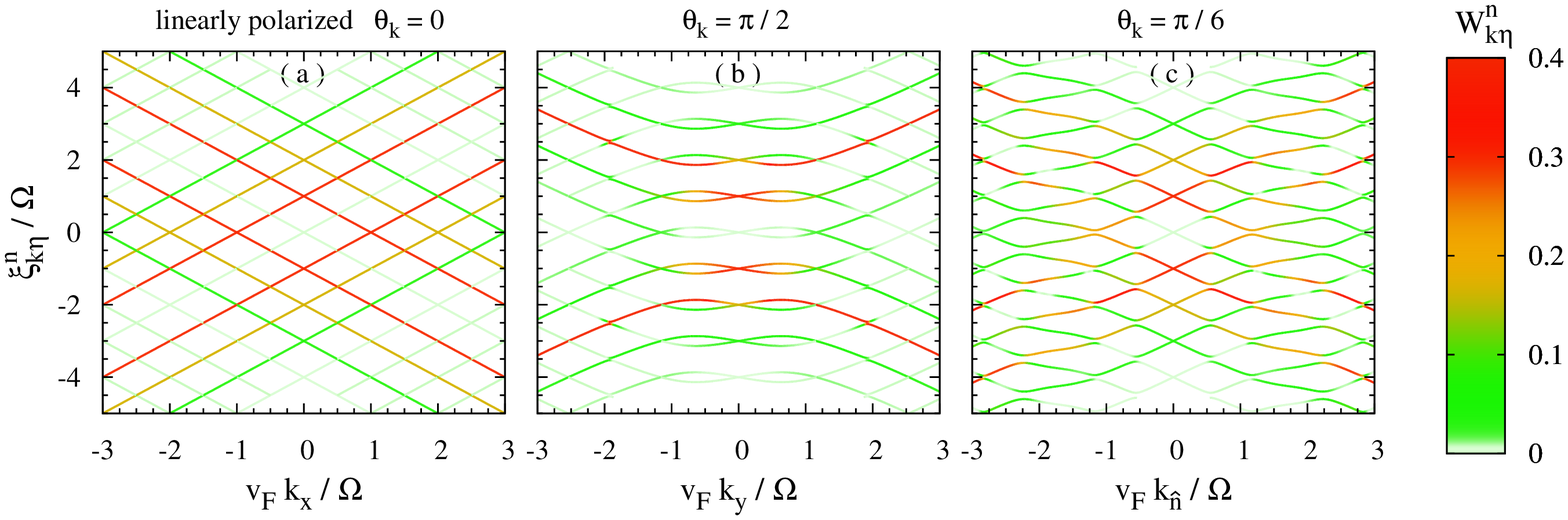}
  \end{center}
  \caption{ \Black{
    (Color online) Linearly polarized THz field with
    $\beta=2.3$. Quasi-energies ${\xi}_{{\bf k}\eta}^{n}$ against the
    normalized momentum for different polar angles 
    $\theta_{\bf k}$. The color coding represents the weight  
    $W_{{\bf k}\eta}^n$ of the corresponding sideband.  
    $k_{\hat{\bf n}}$ in (c) stands for ${\bf k}\cdot{\hat{\bf n}}$
    with $\hat{\bf n}$ being the unit vector at an angle $\pi/6$ with
    respect to the $x$-axis.}
  }
  \label{fig_lin_com}
\end{figure*}

\section{Numerical Results}
In this section, we discuss the numerical results of the energy spectrum, 
DOS and optical conductivity of graphene under an intense THz field.
The typical parameters used in the computation are 
$\Omega=5$~THz, $E_0=15$~kV$/$cm (corresponding to $\beta=2.3$),
$N_e=1.5\times10^{12}$~cm$^{-2}$  
($E_{\rm F}=6\Omega$) and $2.5\times10^{11}$~cm$^{-2}$ ($E_{\rm F}=2.7\Omega$). 
It is noted that our results can be generalized to the other $\Omega$
regime as long as Eq.~(\ref{dis_2}) remains valid,
since the behaviors of the energy spectrum and the DOS are only 
determined by $\beta$ and the behavior of the optical conductivity is only
determined by $\beta$ and $E_{\rm F}/\Omega$.\cite{gamma} 
We also restrict our investigation of optical conductivity in the frequency regime
$\omega_l>\Omega$, since the scattering process is not considered in our model
and the optical conductivity at low frequency is known
to be strongly dependent on the scattering strength.\cite{Orlita_rev,Gusynin,
Ando_02,Neto_06,Stauber_disorder,Stauber_NN,Giuliani_ee,exp_Basov,exp_Nair,exp_APL}

\subsection{Energy spectrum}
\label{energy_spec}
Although the energy spectrum in this system has been
investigated by many works,\cite{Oka_cur,Oka_cur2,Kibis_quantiz,
Zhang_energy,Naumis_pol,Syzranov_08}
a complete investigation on this problem is still lacking. 
In particular, the energy spectrum at large momentum 
has not been well investigated in previous works. 
In this section, we investigate the energy spectrum under
a circularly polarized THz field in a wide range of momentum.
Pronounced quasi-energy gaps appear at small momentum in both cases
with low and high field strengths.
We also discuss the case with a linearly polarized THz field and show
that the energy spectrum becomes anisotropic.

We first concentrate on the case for a circularly polarized THz field
with low field strength (small $\beta$). In
Fig.~\ref{fig_energy_weak}(a), the quasi-energies of  
the sidebands ${\xi}_{{\bf k}\eta}^{n}$  [Eq.~(\ref{energy_side})] for
$\beta=0.4$ are plotted as function of momentum. 
Here the color coding represents the weight $W_{{\bf k}\eta}^n$
[Eq.~(\ref{weight_side})] of the corresponding sideband. 
Since the quasi-energy spectrum is isotropic under a circularly
polarized field, we only present the results with $\theta_{\bf k}=0$.
The most interesting feature seen in this figure is the appearance
of gaps at small momentum in the quasi-energy spectrum, in consistence
with the previous  
investigations.\cite{Oka_cur,Kibis_quantiz,Syzranov_08}  
These gaps appear around the momentums $m\Omega/2v_{\rm F}$ and the
quasi-energies $m\Omega/2+l\Omega$, with $m$ and $l$ being 
integers.
All gaps at the same momentum share the identical magnitude
due to the periodicity of the quasi-energy spectrum.
These quasi-energy gaps can be attributed to the ac Stark splittings
induced by the single-photon/multi-photon
resonances.\cite{Holthaus_Stark, 
Hanggi_98,Faisal_dis2,Hsu_dis2,Martinez_meanenergy} 
The physics is that if a pair of states are coupled by an
electromagnetic field with frequency $\Omega$ and the energy
difference between these two states equals to $m\Omega$, then an ac
Stark splitting appears unless the corresponding transition is forbidden. 
Specifically, in the present case, the gaps around
the momentum satisfying $2v_{\rm F}k=m\Omega$ are
induced by the $m$-photon resonances.\cite{gap_zerok}  

Nevertheless, the behaviour of the quasi-energy spectrum in large
momentum regime becomes very different. 
From Fig.~\ref{fig_energy_weak}(a), one can see that the energy gaps
decrease dramatically with the increase of $k$ 
and finally tend to be closed when the momentum is large enough.
This effect can be understood via 
Eq.~(\ref{gap_approx}) under the rotating-wave approximation 
(the derivation is presented in Appendix~\ref{analy_solution}),
which shows that the quasi-energy gap around the
momentum $m\Omega/2v_{\rm F}$ is determined by the effective coupling
parameter $|y_m(\theta_{\bf k})|$, given by [see also Eq.~(\ref{y_n})] 
\begin{equation}
  |y_m(\theta_{\bf k})|=\frac{1}{\sqrt{2}}|J_{m+1}(\sqrt{2}\beta)
  -J_{m-1}(\sqrt{2}\beta)|,
  \label{y_n_cir}
\end{equation}
with $J_m(\sqrt{2}\beta)$ being the Bessel function. 
Due to the $m$-dependence of $J_m(\sqrt{2}\beta)$, 
the effective coupling parameter $|y_m(\theta_{\bf k})|$ decreases
dramatically with the increase of $m$ when $m$ is large enough. Thus the
quasi-energy gaps become negligible at large momentum. 
Our calculations also show that, associated with the absence of the
gaps, the band-mixing, i.e., the hole (electron) component of the
quasi-electron (quasi-hole) state, becomes negligible, and the
wave function becomes very close to the one without the interband
term of $H_{\rm THz}$. 

In Fig.~\ref{fig_energy_weak}(b), we plot the mean energy 
$\overline{\varepsilon}_{{\bf k}+}$ [Eq.~(\ref{mean_energy})] and the
quasi-energy ${\varepsilon}_{{\bf k}+}$ of the quasi-electron state 
($\overline{\varepsilon}_{{\bf k}+}>0$) as well as the
field-free electron energy for $\beta=0.4$ as function of momentum. 
In order to compare the above three energies directly, we 
choose the quasi-energy ${\varepsilon}_{{\bf k}+}$ according to the 
following rules: 1) the continuity of the quasi-energy is kept as far
as possible;  2) the quasi-energy at large momentum
is closest to the field-free electron energy.
We also plot these selected quasi-energy in
Fig.~\ref{fig_energy_weak}(a) (blue crosses). 
From Fig.~\ref{fig_energy_weak}(b), it is shown that
the mean energy and the quasi-energy are far away from (very close to)
the field-free energy at small (large) momentum, in consistence with
the behaviour of the quasi-energy gap. 
Moreover, it is also seen that the mean energy reaches zero at a
momentum somewhere inside the quasi-energy gap, except the one at zero
momentum. In order to reveal the underlying physics, we plot the
quasi-energy around the gap at $k=0.5\Omega/v_{\rm F}$ and
${\varepsilon}_{{\bf k}\eta}=0.5\Omega$ [the region labelled by the
box in Fig.~\ref{fig_energy_weak}(a)] as well as 
the corresponding mean energy as function of
momentum in the inset of Fig.~\ref{fig_energy_weak}(b).
It shows that there is a crossover point (labelled as
${\bf k}_0$) between the quasi-hole state (blue solid curve) and the
quasi-electron state (red solid curve with crosses) in the
quasi-energy above the gap. Since the quasi-energy varies
continuously with $k$ at this point, the corresponding mean
energy also varies continuously, i.e., 
$\overline{\varepsilon}_{{\bf k}_0+}=
\overline{\varepsilon}_{{\bf k}_0-}$.
Also, from Eqs.~(\ref{symmetry_1}) and (\ref{symmetry_2}), one can see
that $\overline{\varepsilon}_{{\bf k}_0+}=
-\overline{\varepsilon}_{{\bf k}_0-}$.
Consequently, the mean energies at this point must be zero.
The finite mean energy at $k=0$ is based on the similar reason. 
As shown in Fig.~\ref{fig_energy_weak}(a), the quasi-energy above the
gap at zero momentum always belongs to the quasi-electron state [blue
crosses in Fig.~\ref{fig_energy_weak}(a)], hence no crossover 
appears.

Now we turn to the case for a circularly polarized THz field with
high field strength (Fig.~\ref{fig_energy_strong}). 
From Fig.~\ref{fig_energy_strong}(a), it is seen that pronounced
quasi-energy gaps appear in a wider range of $k$.
However, unlike the previous case, the
momentums of the gaps markedly deviate from $m\Omega/(2v_{\rm F})$. 
This is because the effect of the rapidly varying terms 
in the resonance equations [Eqs.~(\ref{equ_a1_m}) and
(\ref{equ_a2_m})] cannot be neglected for strong field, 
i.e., the rotating-wave approximation is not valid. 
The joint effect of the terms with different oscillating 
frequencies leads to the complicated behaviour of the quasi-energy
spectrum. However, the quasi-energies where the gaps appear are still 
around $l\Omega/2$, determined by the symmetry of the
quasi-energies of the sidebands [Eq.~(\ref{symmetry_3})]. 
Figure~\ref{fig_energy_strong} also shows that the quasi-energy gaps
become extremely small and the quasi-energy and mean energy become
very close to the field-free energy at large momentum. 
The physics is similar to the weak-field case.
When the photon number involved in the resonance becomes too large,
the effective resonant coupling becomes extremely weak.
Thus the influence of the THz field on the energy spectrum becomes
negligible.

Finally we address the case for a linearly polarized THz field
(Fig.~\ref{fig_lin_com}). 
Here we only present the results with high field strength as they
are sufficient for showing the main features in this case.
Recall that the linearly polarized THz field is set along the $x$-axis
throughout this paper, thus $\theta_{\bf k}$ equals to the 
angle between the momentum and the THz field. 
From Fig.~\ref{fig_lin_com}, it is seen that quasi-energy gaps are
closed at zero momentum. 
This effect can be understood from the exact analytical solution
\begin{eqnarray}
  |\Phi_{{\bf k}=0\,\pm}(t) \rangle &=& \sum_{n} J_{\pm n}(\beta) e^{in\Omega t}
  (1,\;\pm1)^{\rm T}.
  \label{Phi_k0}
\end{eqnarray}
Clearly, the quasi-energy is zero at zero momentum, thus the
gap disappears. 
Another interesting feature is the anisotropy of the
quasi-energy spectrum.\cite{Syzranov_08}  
For momentum along the $x$-axis [Fig.~\ref{fig_lin_com}(a)], all gaps
are closed, since the interband term of ${H}_{\rm THz}$
[Eq.~(\ref{H_THz_intra})] becomes zero and hence the quasi-energy is
exactly the same as the field-free energy, as shown in
Eqs.~(\ref{Psi_0+}) and (\ref{Psi_0-}) in Appendix~\ref{analy_solution}.
For momentum along the $y$-axis [Fig.~\ref{fig_lin_com}(b)],
all gaps except the ones around $k=0.5\Omega/v_{\rm F}$ are
effectively closed. 
This is because the intraband term of ${H}_{\rm THz}$
[Eq.~(\ref{H_THz_intra})] becomes zero, 
and thus the gaps from the multi-photon resonances become extremely
small, as shown in Fig.~\ref{fig_gap} in Appendix~\ref{analy_solution}.
For other polar angle, e.g., $\theta_{\bf k}=\pi/6$
[Fig.~\ref{fig_lin_com}(c)], pronounced quasi-energy 
gaps appear not only around $k=0.5\Omega/v_{\rm F}$ but also around
the other resonant points at small momentum, similar to
the case with a circularly polarized THz field.

\begin{figure}[tbp]
  \begin{center}
    \includegraphics[width=6.2cm]{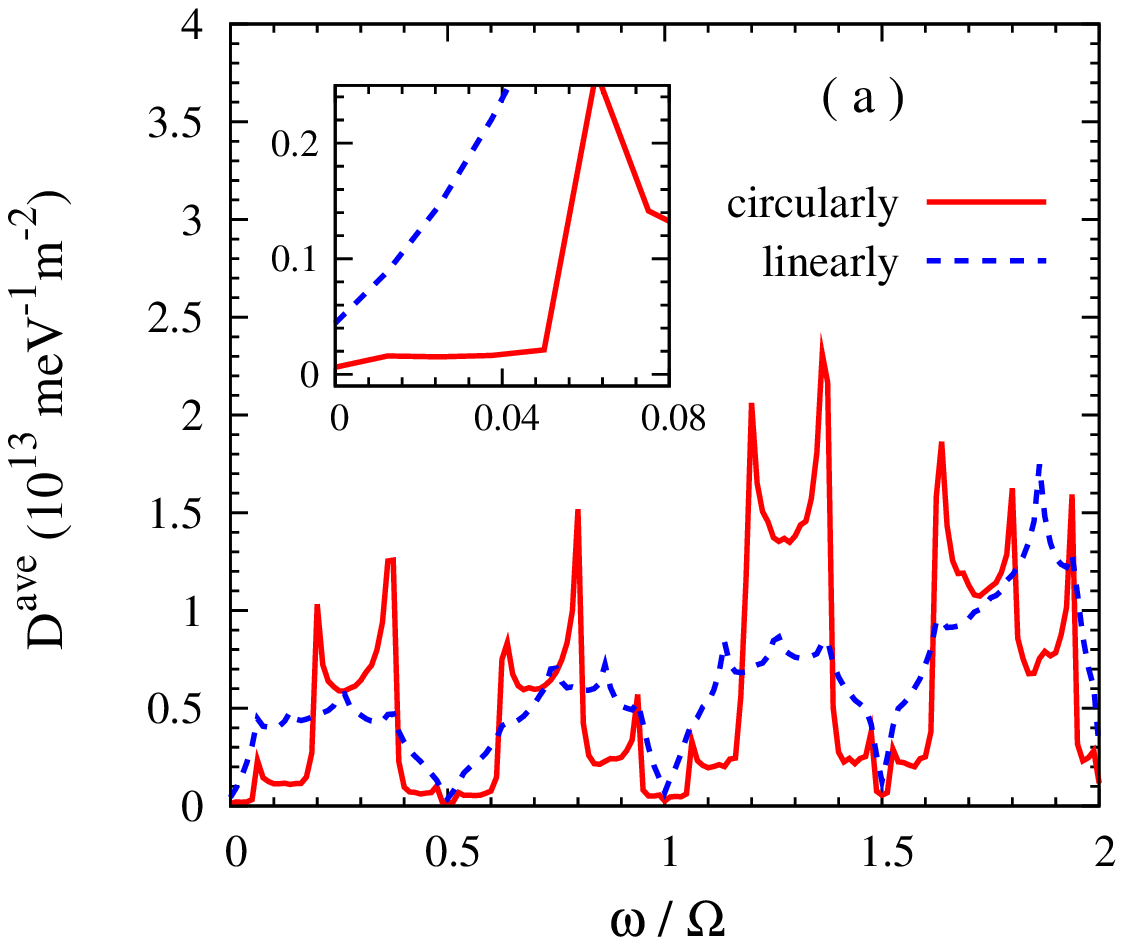}
  \end{center}
  \begin{center}
    \includegraphics[width=6.2cm]{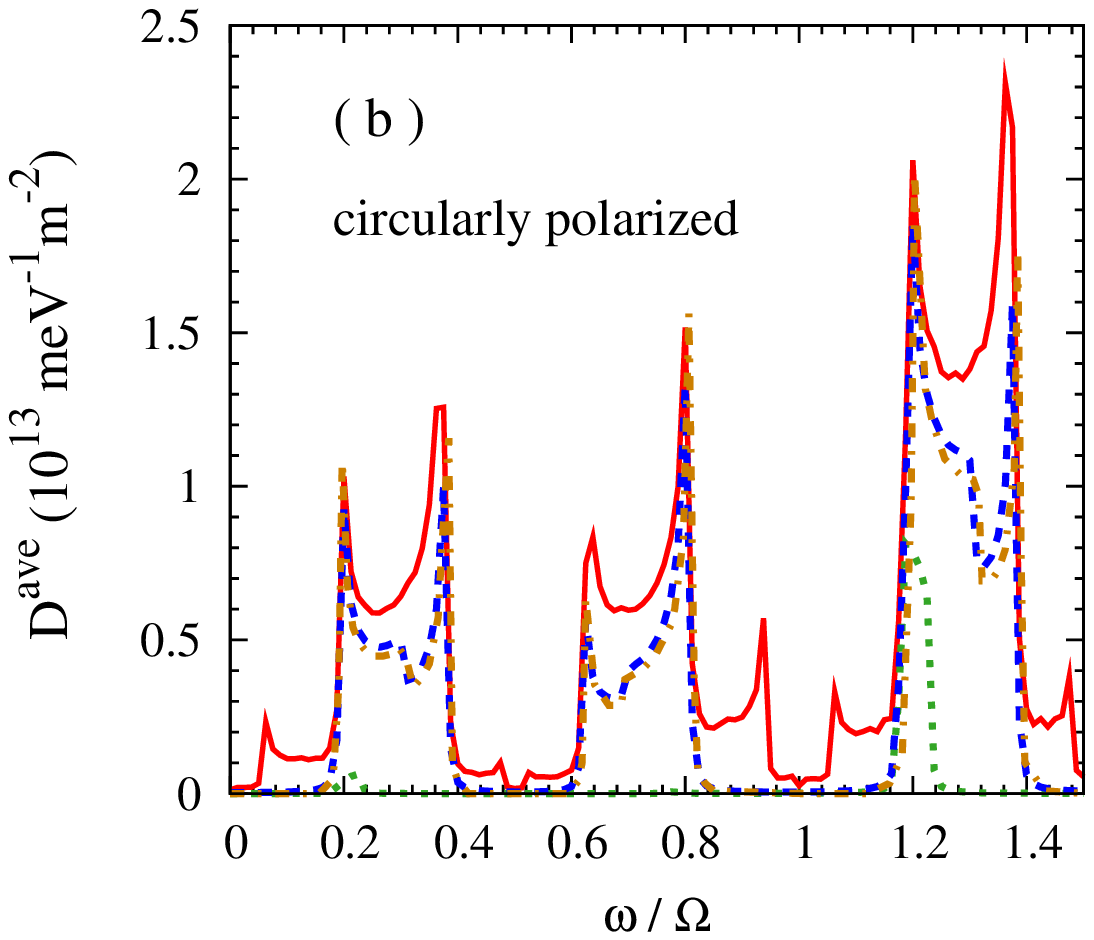}
  \end{center}
  \caption{ (Color online) (a) Time-averaged DOS versus $\omega$
     under circularly (red solid curve) and linearly (blue dashed
     curve) polarized THz fields for $\beta=2.3$.   
     The region close to $\omega\sim0$ is enlarged in the inset. 
     (b) Time-averaged DOS versus $\omega$ under a circularly
     polarized THz field for $\beta=2.3$ with all relevant states (red
     solid curve), limited in the momentum regimes 
     $v_{\rm F}k<2\Omega$ (blue dashed curve) and 
     $v_{\rm F}k<0.5\Omega$ (green dotted curve).
     The yellow chain curve represents the DOS from
     Ref.~\onlinecite{Oka_cur}. 
  }
  \label{fig_DOS}
\end{figure}

\subsection{DOS}
Then we turn to investigate the DOS. By using the spectral function
Eq.~(\ref{A_omega}), one obtains the time-averaged DOS
\begin{eqnarray}
  \nonumber
  {D}^{\rm ave}(\omega) &=& \frac{g_sg_v}{2\pi T_0} \int_0^{T_0}dT \; 
  \sum_{{\bf k}\eta} {\rm Tr}{A}_{{\bf k}\eta}(T,\omega)\\
  \nonumber
  &=&g_sg_v\sum_{{\bf k}\eta n}
  \langle \phi_{{\bf k}\eta}^{n}| \phi_{{\bf k}\eta}^{n} \rangle
  \delta(\omega-\varepsilon_{{\bf k}\eta} +n\Omega)\\
  &=&g_sg_v\sum_{{\bf k}\alpha}
  \langle \psi_{{\bf k}\alpha}^0| \psi_{{\bf k}\alpha}^0 \rangle
  \delta(\omega-\mathcal{E}_{{\bf k}\alpha}).
  \label{DOS_ave}
\end{eqnarray}
The last equality is derived from Eqs.~(\ref{relation_1}) and
(\ref{relation_2}). 
It is noted that the above equation is in the same form as 
the one reported by Oka and Aoki.\cite{Oka_cur}  
In Fig.~\ref{fig_DOS}(a), we plot the time-averaged DOS under 
circularly and linearly polarized THz fields for $\beta=2.3$.
(Note that $\beta=2.3$ corresponds to $F/\Omega=\beta/\sqrt{2}=1.6$ in
Ref.~\onlinecite{Oka_cur}.\cite{other_def})
The region close to $\omega\sim0$ is enlarged in the inset.
Here we only show the DOS in the positive frequency
regime (electron regime), since the DOS in the negative regime (hole
regime) is symmetrical to the positive one. 

We first focus on the case for a circularly polarized THz field 
[red solid curve in Fig.~\ref{fig_DOS}(a)].
It is seen that the gaps in the DOS are effectively {\em closed}, in
contrast to the previous report by Oka and Aoki.\cite{Oka_cur}
The underlying physics is as follows: as shown in
Fig.~\ref{fig_energy_strong}(a), the quasi-energy gaps are closed when 
the momentum is large enough. 
The contribution from these states closes the gaps in the DOS.
We also plot the DOS with all related states (red solid curve),
limited in the momentum regimes $v_{\rm F}k<2\Omega$ (blue dashed
curve) and $v_{\rm F}k<0.5\Omega$ (green dotted curve) in
Fig.~\ref{fig_DOS}(b).  
It is shown that the DOS with $v_{\rm F}k<2\Omega$ 
(blue dashed curve) is almost the same as 
the one shown in Ref.~\onlinecite{Oka_cur} (yellow chain curve).
This indicates that the contribution from large momentum
is not properly counted in the previous investigation.\cite{Oka_cur}
In addition, it is seen that sharp peaks appear in the
DOS. This effect originates from the van Hoff singularities 
which appear at the momentum of all quasi-energy 
gaps due to the isotropic quasi-energy spectrum.\cite{sing_zerok}

Then we turn to the case of a linearly polarized THz field [blue
dashed curve in Fig.~\ref{fig_DOS}(a)].  
It is also shown that the gaps are absent. However
the physics is different here. As shown in Fig.~\ref{fig_lin_com}(a),
the quasi-energy gaps disappear for the momentum along the
$x$-axis, even at small $k$, and the gaps due to the multi-photon
resonances become negligible for momentum 
along the $y$-axis. Therefore there is no gap in the DOS. 
Moreover, the peaks from the van Hoff singularities are less
pronounced due to the different nature of the van Hoff singularities 
here associated with anisotropic quasi-energy spectrum.\cite{van_Hoff}

\begin{figure}[tbp]
 \begin{center}
    \includegraphics[width=6.5cm]{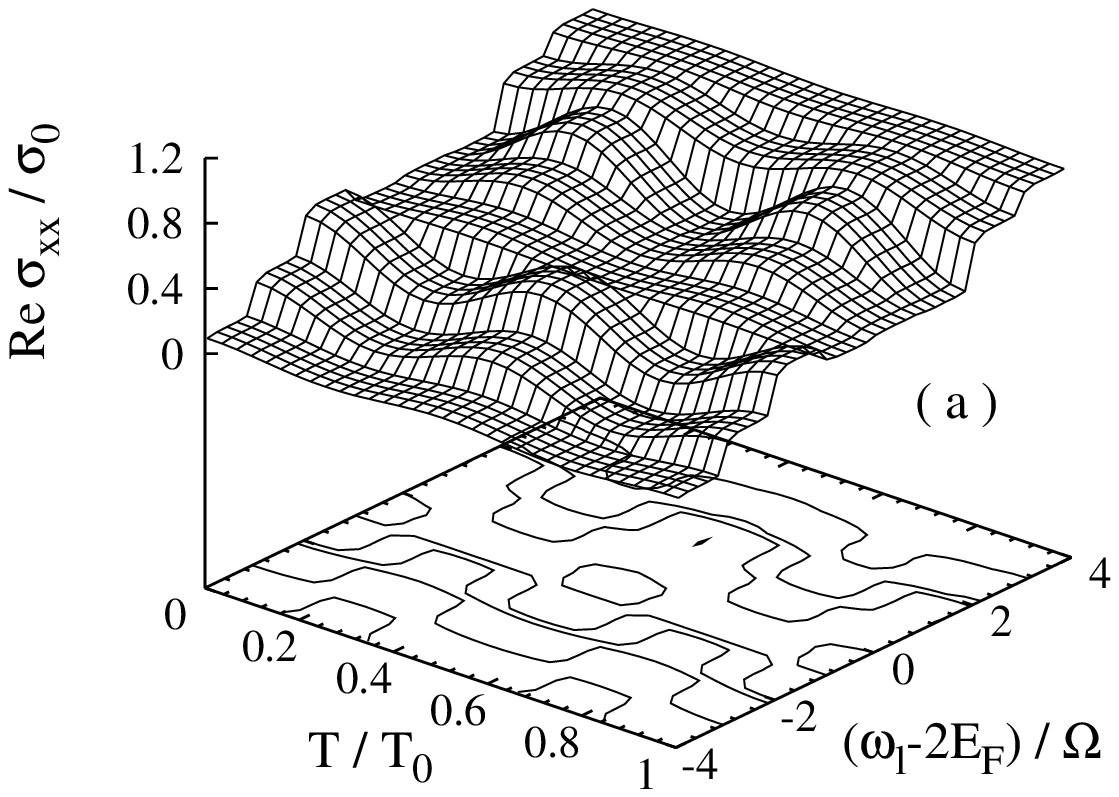}
  \end{center}
  \begin{center}
    \includegraphics[width=6.cm]{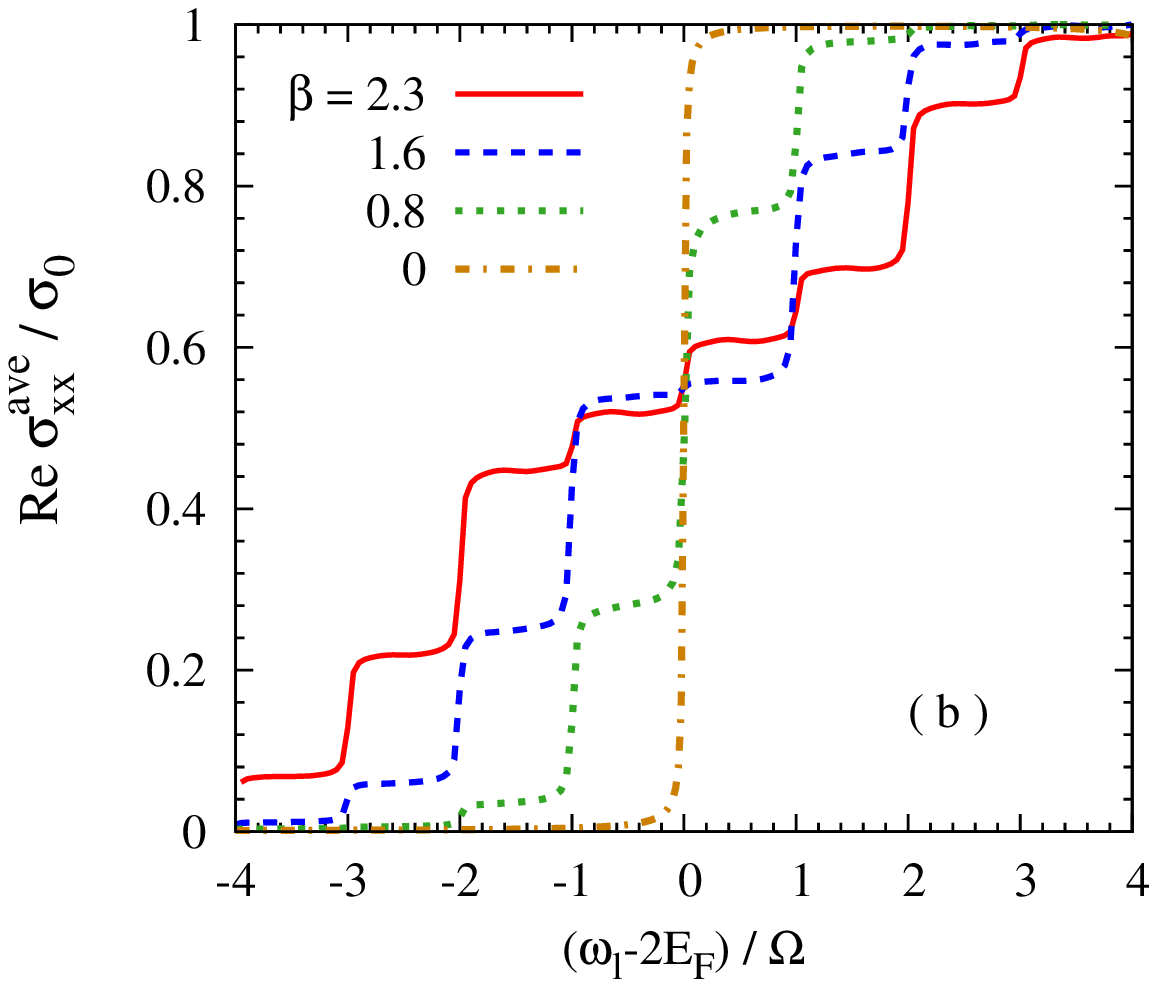}
  \end{center}
  \caption{ (Color online)  Circularly polarized THz field. 
    (a) Time-dependent optical conductivity as function of the optical
    frequency and  time with $\beta=2.3$. 
    (b) Time-averaged optical conductivity as function of the optical
    frequency with different field strengths.
    $E_{\rm F}=6\Omega$ in both figures.  
  }
  \label{fig_cir_hEf}
\end{figure}

\begin{figure}[tbp]
  \begin{center}
    \includegraphics[width=6.5cm]{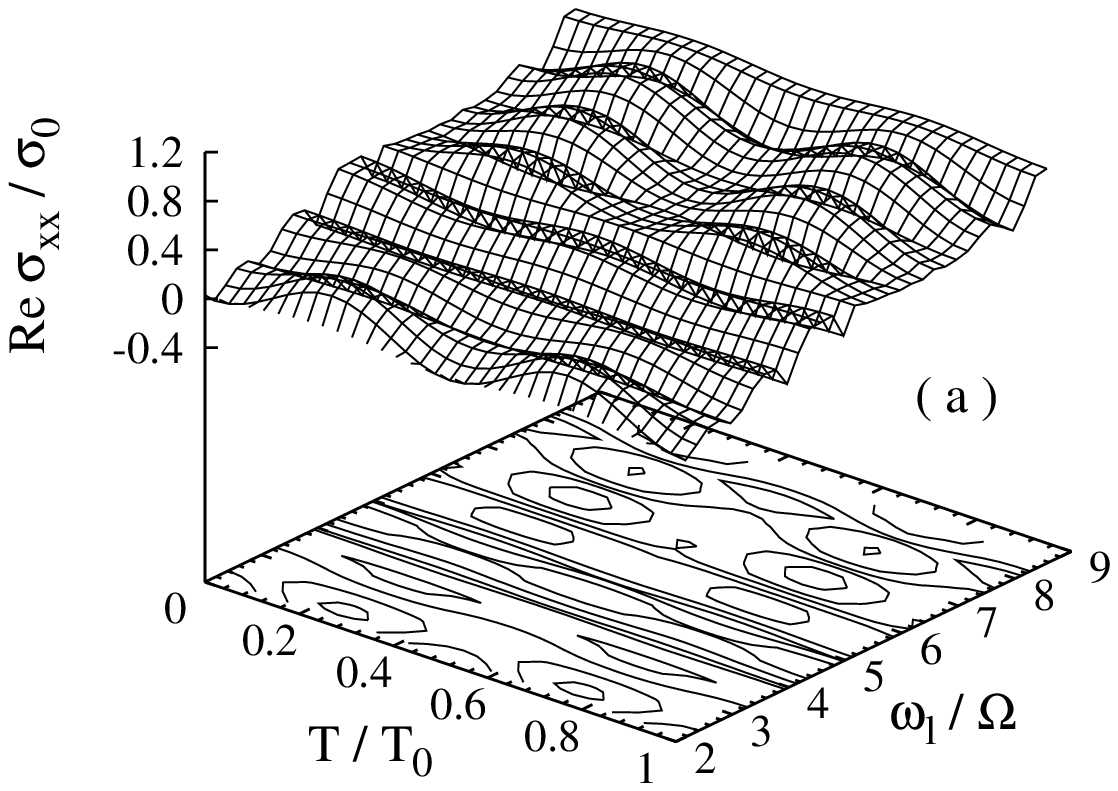}
  \end{center}
  \begin{center}
    \includegraphics[width=6.cm]{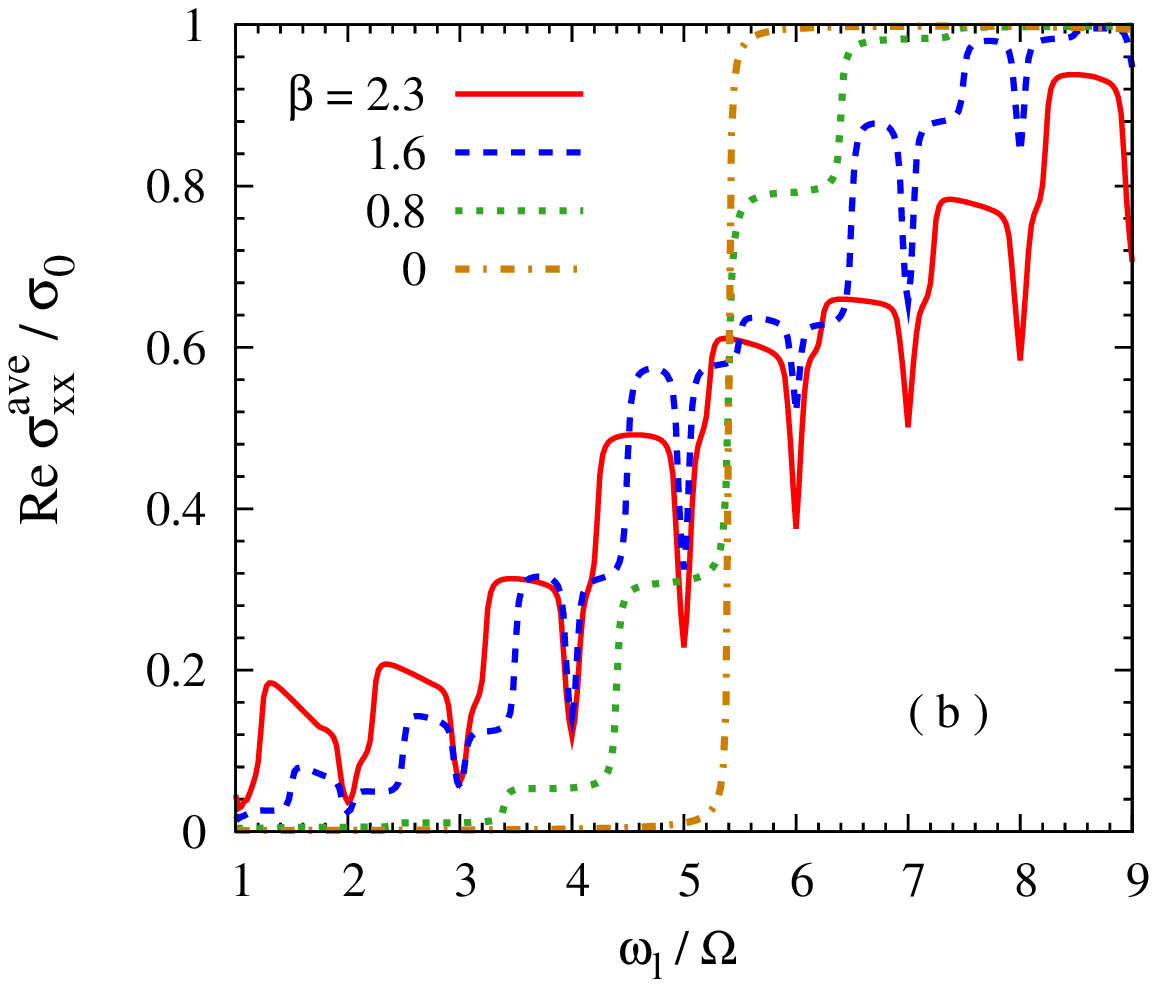}
  \end{center}
  \begin{center}
    \includegraphics[width=6.2cm]{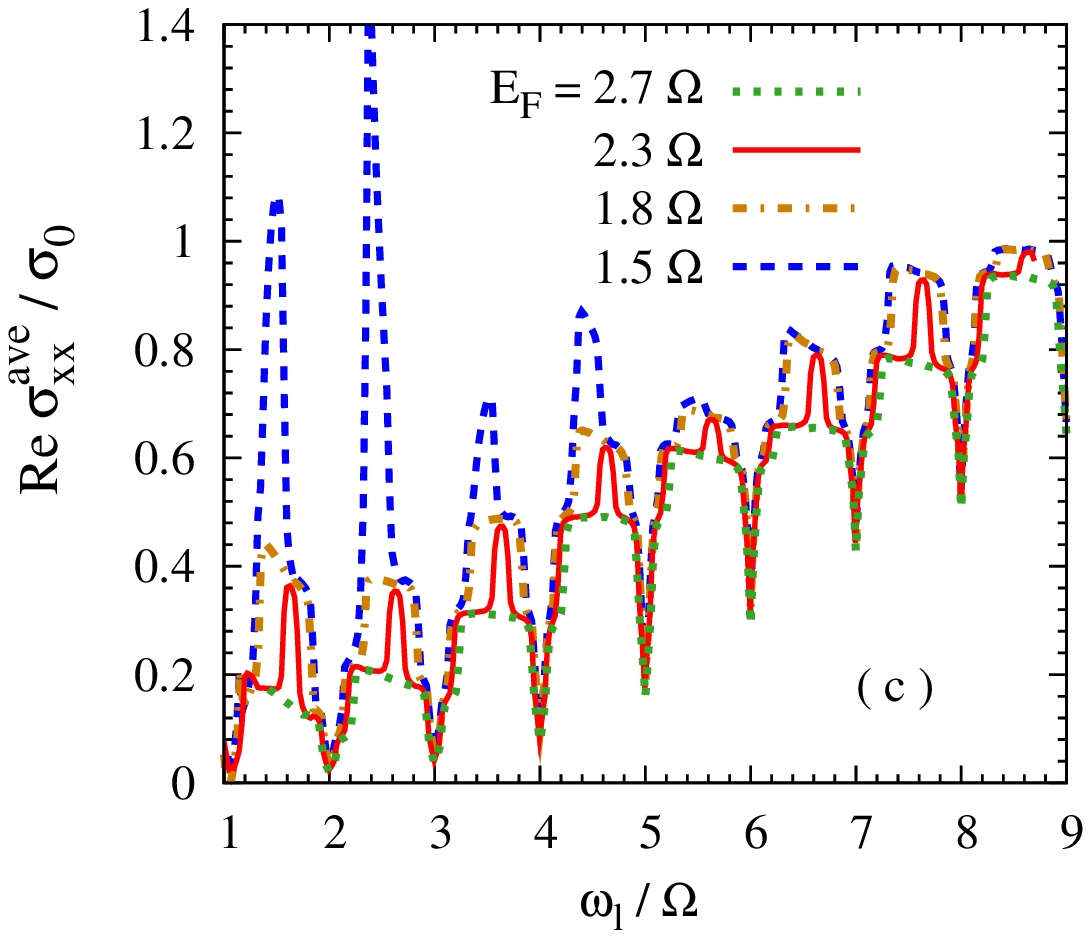}
  \end{center}
  \caption{ (Color online) Circularly polarized THz field.
    (a) Time-dependent optical conductivity as function of the
    optical frequency and time with $\beta=2.3$ and
    $E_{\rm F}=2.7\Omega$. 
    (b) Time-averaged optical conductivity as function of the
    optical frequency with different field strengths.
    $E_{\rm F}=2.7\Omega$.  
    (c) Time-averaged optical conductivity as function of the
    optical frequency with $\beta=2.3$ for different Fermi energies. 
  }
  \label{fig_cir_lEf}
\end{figure}

\subsection{Optical conductivity}
\subsubsection{Under a circularly polarized THz field}
In this subsection we discuss the optical conductivity under a
circularly polarized THz field. 
Without loss of generality, we restrict ourselves to the $n$-type 
case, i.e., $E_{\rm F}>0$. 
Thus only the quasi-electron states with
$\overline{\varepsilon}_{{\bf k}\eta}>E_{\rm F}$ are not occupied and the
the corresponding interband transitions are allowed.

We first focus on the case with high Fermi energy.
The time-dependent and time-averaged optical conductivities as function
of the optical frequency $\omega_l$ are plotted
in Figs.~\ref{fig_cir_hEf}(a) and (b), respectively.
Here $\sigma_0=\frac{e^2}{4\hbar}$
is used as the unit of the optical conductivity. 
It is shown that the optical conductivity 
exhibits a multi-step-like structure at $\omega_l\sim 2E_{\rm F}$, in
contrast to the single-step-like behaviour in the field-free case
[yellow chain curve in
Fig.~\ref{fig_cir_hEf}(b)].\cite{Orlita_rev,Gusynin,Ando_02,Neto_06,
Stauber_disorder,Stauber_NN,Giuliani_ee,exp_Basov,exp_Nair,exp_APL}
Similar to the DFK effect in semiconductors,\cite{Jauho_DFK}
this effect is from the sideband-modulated optical transition,
i.e., $n\Omega$ in the delta functions in Eqs.~(\ref{opt_T}) and
(\ref{opt_ave}). 
It is also seen that the number of ``step'' increases with the
increase of the field strength. This can be understood by
noticing that the weight of the sideband is distributed in a
wider range of frequency for stronger field. 

Figure~\ref{fig_cir_hEf} also shows that the optical
conductivity varies mildly with the increase of $\omega_l$ 
in each ``step'', which is quite different
from the DFK effect in semiconductors\cite{Jauho_DFK}
where the optical absorption is strongly dependent on $\omega_l$ 
in each ``step''.
This behavior can be understood as follows.
Since the effect from the interband term of ${H}_{\rm THz}$ becomes
negligible at  large momentum, the optical conductivity at high Fermi
energy can be approximately described by Eq.~(\ref{ana_fin_dis2})
(the derivation is presented in Appendix~\ref{analy_opt}),
which indicates that the frequency dependence of the optical
conductivity is only from the factor  
$1-N{\Omega}/{\omega_l}$ in the optical
  transition between the sidebands with the energy difference 
  $2\overline{\epsilon}_{{\bf k}+}+N\Omega$. This factor originates
  from the linear dispersion of graphene.
It is also noted that the optical conductivity is dominated by 
the optical transitions with small $|N|$,
and the pronounced ``steps'' only appear at $\omega_l\sim 2E_{\rm F}$.
Thus the frequency dependence of the optical conductivity
becomes very weak in each ``step'' for high Fermi energy. 

\begin{figure}[tbp]
  \begin{center}
    \includegraphics[width=5.8cm]{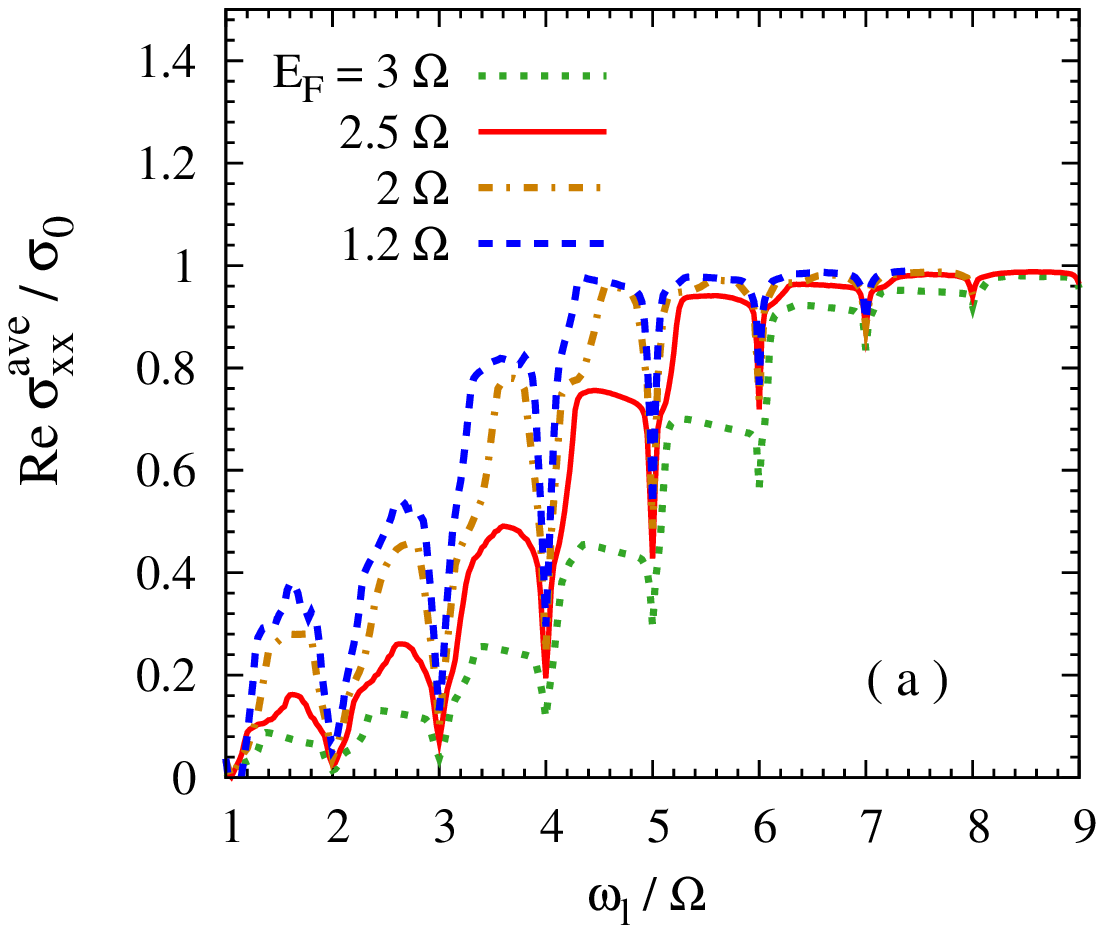}
  \end{center}
  \begin{center}
    \includegraphics[width=5.8cm]{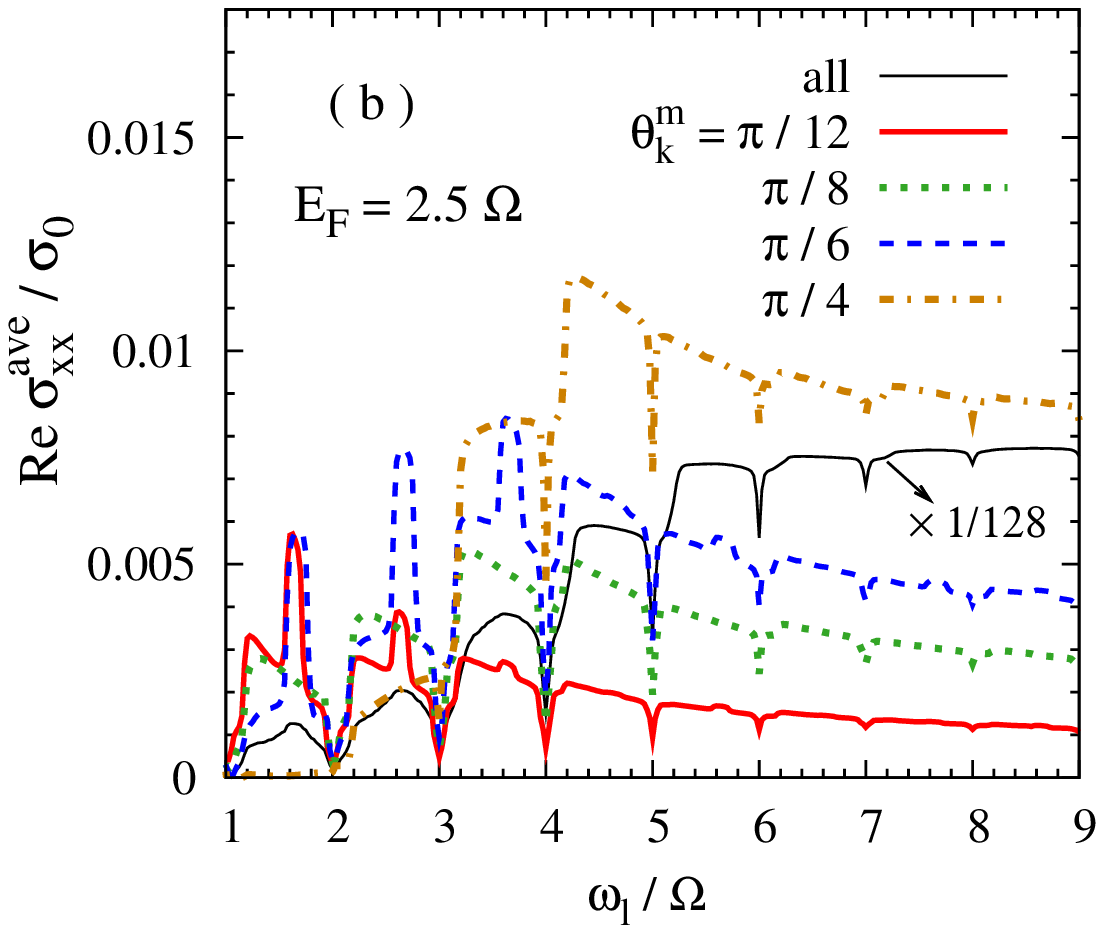}
  \end{center}
  \begin{center}
    \includegraphics[width=5.8cm]{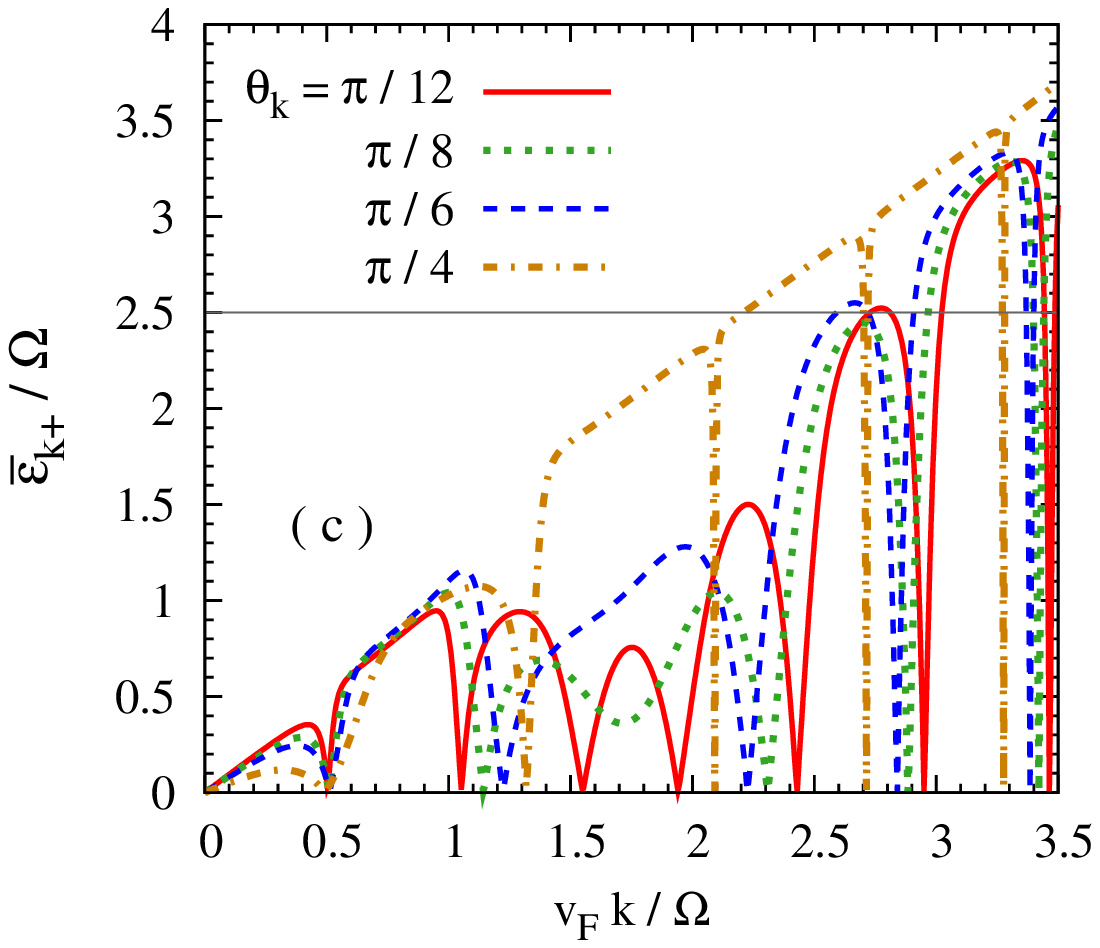}
  \end{center}
  \caption{ (Color online) Linearly polarized THz field with $\beta=2.3$.
    (a) Time-averaged optical conductivity as function of the optical frequency
    for different Fermi energies. (b) Time-averaged optical
    conductivities versus the optical frequency from the states in
    different polar angle regions 
    $[\theta^m_{\bf k}-\pi/128,\theta^m_{\bf k}+\pi/128)$ as well as
    the one with all polar angles integrated. $E_{\rm F}=2.5\Omega$. 
    (c) Mean energies of the quasi-electron states with different
    polar angles $\theta_{\bf k}$ against the normalized momentum. 
    The thin black line indicates the mean energy being $2.5\Omega$.
  }
  \label{fig_lin}
\end{figure}

Then we turn to the case with low Fermi energy. The time-dependent and
time-averaged optical conductivities are plotted as function of the 
optical frequency in the cases with $E_{\rm F}=2.7\Omega$ in 
Figs.~\ref{fig_cir_lEf}(a) and (b), respectively.
It is seen that dips appear around the frequencies satisfying 
$l\Omega$ when the applied THz field is strong enough. 
This is because the states at small momentum, where the
quasi-energy gaps become pronounced as shown in
Fig.~\ref{fig_energy_strong}(b), can contribute to the optical
conductivity in this case. 
More interesting features are presented in the Fermi energy dependence
of optical conductivity [Fig.~\ref{fig_cir_lEf}(c)].
It is shown that peaks appear in the middle of 
the ``steps'' for $E_{\rm F}=2.3\Omega$ (red solid curve). 
The scenario is as follows. From Fig.~\ref{fig_energy_strong}(b), 
it is seen that the maximum of the mean energy $\overline{\varepsilon}_{\bf k+}$ 
at $2.5\Omega/v_{\rm F}$ is slightly higher than the Fermi energy
$2.3\Omega$. Consequently, in this momentum regime only the
quasi-electron states around $2.5\Omega/v_{\rm F}$ can contribute to
the optical conductivity and hence induces the peaks in the middle of
the ``steps''. When the Fermi energy decreases, more and more states
in this momentum regime can contribute to the optical
conductivity. Consequently the peaks become wider and wider, and
finally become the new ``steps'', as 
shown in the case with $E_{\rm F}=1.8\Omega$ (yellow chain curve).
Moreover, when the Fermi energy decreases to $1.5\Omega$ (blue dashed
curve), sharp peaks appear in the optical conductivity. 
This originates from the contribution around $k=0$, in which  
$\overline{\varepsilon}_{\bf k+}$ is slightly higher than
$1.5\Omega$ as shown in Fig.~\ref{fig_energy_strong}(b).
One also notes that the van Hoff singularities at nonzero
momentum have no effect on
the optical conductivity with finite Fermi energy,
since the mean energy is close to zero at the momentum of
the quasi-energy gaps, as shown in Fig.~\ref{fig_energy_strong}(b).

\subsubsection{Under a linearly polarized THz field}
We also plot the time-averaged optical conductivity as function of the
optical frequency under a linearly polarized THz field in
Fig.~\ref{fig_lin}(a).  
The behavior in this case is simpler than that under a circularly
polarized THz field. The multi-step-like behavior and the dips around
$l \Omega$ still appear. Nevertheless, the peaks in the middle of 
the ``steps'', which still appear in the optical conductivity at low
frequency for $E_{\rm F}=2.5\Omega$ and $1.2\Omega$, are much less
pronounced than the ones under a circularly polarized THz field.
In order to reveal the underlying physics, we plot the time-averaged
optical conductivity from the states with different polar angles
as function of the optical frequency for $E_{\rm F}=2.5\Omega$
in Fig.~\ref{fig_lin}(b). The corresponding mean energies are plotted
against the normalized momentum in Fig.~\ref{fig_lin}(c).  
As mentioned above, the peaks in the middle of the ``steps'' only appear in the
situation satisfying the criteria that a local maximum of the mean energy
$\overline{\varepsilon}_{\bf k+}$ is slightly higher than the Fermi energy.  
From Fig.~\ref{fig_lin}(c), one can see that the 
above criteria is satisfied for the states
with $\theta_{\bf k}=\pi/12$ and $\pi/6$, thus pronounced peaks appear
in the optical conductivity from the states with these angles [red
solid and blue dashed curves in Fig.~\ref{fig_lin}(b)] 
and the corresponding angles with the same quasi-energy.\cite{angle_symmetry} 
However, this criteria cannot be satisfied for the states
with the other polar angles,
e.g., $\theta_{\bf k}=\pi/8$ and $\theta_{\bf k}=\pi/4$,
so peaks are absent in the corresponding optical conductivity [green dotted
and yellow chain curves in Fig.~\ref{fig_lin}(b)].
Therefore, after the summation of the contribution from all polar angles,
the peaks in the middle of the ``steps'' become much less pronounced.

\subsubsection{Comparison of the optical conductivities calculated with
 the mean-energy-determined distribution and the projected  distribution}
In this subsection, we compare the optical conductivities calculated 
with the distribution determined by the mean energy and the projected
distribution. As mentioned above, the projected distribution used in
Ref.~\onlinecite{Oka_opt} is described by (see also
Appendix~\ref{modified_oka}) 
\begin{equation}
  f_{{\bf k}\eta}^{\rm ini}=\sum\limits_{\nu l}n_{\rm F}(E_{{\bf k}\nu})
  |\langle\zeta_{{\bf k}\nu}|\phi_{{\bf k}\eta}^{l}\rangle|^2.
  \label{dis_ini}
\end{equation}
Recall that $E_{{\bf k}\nu}$ and $\zeta_{{\bf k}\nu}$ are the
eigenvalue and eigenvector of $\hat{H}_0$.
From Eq.~(\ref{dis_ini}), one can recognize the main features of the
projected distribution at zero temperature: 
at $k<k_{\rm F}=E_{\rm F}/v_{\rm F}$, both the field-free electron and
hole states are occupied, thus the corresponding quasi-electron
and quasi-hole states are both occupied; 
at $k>k_{\rm F}$, only the hole state is occupied, so the
distribution of the Floquet state is determined by 
the hole component of this state. 
Therefore one can see that only the states with $k>k_{\rm F}$ can
contribute to the optical conductivity, and the contribution
decreases with the increase of the band-mixing.

\begin{figure}[tbp]
  \begin{center}
    \includegraphics[width=6.2cm]{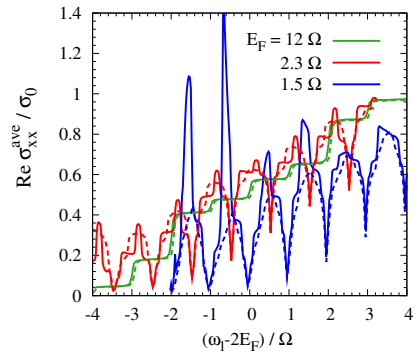}
  \end{center}
  \caption{ (Color online) Time-averaged optical conductivities from the
    mean-energy-determined distribution (solid curves) and the projected
    distribution (dashed curves) are plotted as function of the
    optical frequency under a circularly polarized THz field with
    $\beta=2.3$ for different Fermi energies. 
  }
  \label{fig_com}
\end{figure}

The time-averaged optical conductivities from the
mean-energy-determined distribution and the projected distribution are
plotted as function of the optical frequency in Fig.~\ref{fig_com}.
It is seen that in the case with high Fermi energy, the results from 
both distributions are almost the same. 
This is because the optical conductivity in this situation is
determined by the contribution from the states at large momentum,
where the band-mixing is negligible due to the marginal effective
resonant coupling.  
Consequently these two optical conductivities are very close to
the approximate formula Eq.~(\ref{ana_fin_dis2}) in
Appendix~\ref{analy_opt}. 
However, at low Fermi energy, 
it is seen that although the dips due to the quasi-energy gaps also
appear in the optical conductivity from the projected distribution,
the peaks in the middle of the ``steps'' and the peaks
from the states around zero momentum are absent. 
The absence of both effects is because the quasi-electron
states with small $k$ are occupied in the projected distribution.  
Thus the corresponding optical transitions are blocked.
In addition, it is also shown that the van Hoff singularities do not
affect on the optical conductivity from the projected distribution,
since the contribution from the states around the momentum of the
quasi-energy gaps is extremely small due to the strong band-mixing. 

\section{Summary and discussion}
In summary, we have performed a theoretical investigation on the
optical responses of graphene in the presence of intense circularly and
linearly polarized THz fields via the Floquet theory. 
We examine the energy spectrum and DOS. 
It is found that gaps appear at small momentum in the quasi-energy 
spectrum, in consistence with the previous 
investigations.\cite{Oka_cur,Kibis_quantiz,Syzranov_08}  
These gaps can be attributed to the ac Stark splittings induced by the
single-photon/multi-photon resonances. Nevertheless, in large
momentum regime, where the energy spectrum has not been well
investigated in previous works, we find that the quasi-energy gaps
decrease dramatically with the increase of momentum and finally tend 
to be closed when the momentum is large enough. 
Consequently, taking account of the contribution from the states at
large momentum, the gaps in the DOS are effectively closed, in
contrast to the prediction by Oka and Aoki.\cite{Oka_cur}

We also investigate the optical conductivity from the
mean-energy-determined distribution for different field
strengths and Fermi energies.
These results reveal the main features of the DFK effect in graphene.
In the case with high Fermi energy, we discover that the optical
conductivity presents a multi-step-like 
behavior around the optical frequency $\omega_l$ twice of the Fermi
energy $E_F$, in contrast to the single-step-like behaviour in the
field-free case.
This effect is from the sideband-modulated optical transition,
similar to the DFK effect in semiconductors.
We also find that the optical conductivity varies mildly with the
increase of $\omega_l$ in each ``step'', which is quite different from
the DFK effect in semiconductors. 
This behaviour is due to the linear dispersion of graphene
and the absence of the band-mixing at large momentum.

The behaviour of the optical conductivity becomes more interesting in
the case with low Fermi energy. We discover that dips appear at
frequencies being the integer numbers of the applied 
THz field frequency $\Omega$, due to the quasi-energy gaps at
small momentums. In the case with a circularly polarized THz field,
it is found that peaks appear in the middle of the ``steps''
when the Fermi energy is slightly lower than a local maximum of the mean
energy. This kind of peaks become much less pronounced in the case with a
linearly polarized field, owing to the anisotropic energy spectrum.
Another interesting finding in the case with a circularly polarized
field is that the contribution from the states around zero momentum
can induce sharp peaks in the optical conductivity when the Fermi
energy is lower than the mean energy at $k=0$. 

Finally, we address the distribution function of the Floquet states.
Our calculations are based on the ansatz that the distribution
function of the Floquet states is determined by the mean energy, 
following the works in the literature.\cite{Gupta_dis2,Faisal_dis2,Hsu_dis2}
 The projected distribution
function is also adopted in the literature.\cite{Oka_opt}
It is noted that the multi-step-like behavior and the dips around 
frequencies $l\Omega$ exist in optical conductivities from
both distributions. This indicates that these two effects do not 
depend on the details of the distribution function and thus are
expected to be observed in the 
optical absorption measurements subject to intense THz fields. 
Nevertheless, the peaks in the middle of the ``steps'' and the peaks
from the states around zero momentum appear only in the optical
conductivity from the mean-energy-determined distribution. 
By performing experimental investigation on these peaks in 
the optical conductivity, one can distinguish which distribution function of the
Floquet state is closest to the genuine one. One may also solve 
the kinetic equations with all the scattering explicitly included\cite{Wu_rev}
to determine the distribution function.

\begin{acknowledgments}
  This work was supported by the National Natural Science Foundation of China
  under Grant No.\ 10725417 and the Knowledge Innovation Project of Chinese
  Academy of Sciences. One of the authors (M.W.W.) acknowledges discussions
  with M. Gonokami. Y. Z. gratefully thanks J. H. Jiang and K. Shen for
  their help in this work.
\end{acknowledgments}

\begin{appendix}
\section{Optical conductivity from extend Kubo formula}
\label{modified_oka}
We write Eq.~(3) in Ref.~\onlinecite{Oka_opt} into the form
\begin{eqnarray}
  \nonumber
  {\rm Re}{\sigma}^{\rm ave}_{xx}(\omega_l)&=&\frac{4\pi}{\omega_l}
  \sum_{{\bf k}\alpha\beta \atop nm}
  \langle\psi_{{\bf k}\alpha}^n| \hat{j}^x_{{\bf k}}
  |\psi_{{\bf k}\beta}^n\rangle
  \langle\psi_{{\bf k}\beta}^m| \hat{j}^x_{{\bf k}}
  |\psi_{{\bf k}\alpha}^m\rangle\\ &&
  \mbox{}\times (\tilde{f}_{{\bf k}\alpha}-\tilde{f}_{{\bf k}\beta})
  \delta(\mathcal{E}_{{\bf k}\alpha}-\mathcal{E}_{{\bf k}\beta}+\omega_l)
\end{eqnarray}
with $\tilde{f}_{{\bf k}\alpha}=\sum\limits_\nu|\langle\zeta_{{\bf k}\nu}|
\psi_{{\bf k}\alpha}^0\rangle|^2 n_{\rm F}(E_{{\bf k}\nu})$. 
Recall that $E_{{\bf k}\nu}$ and $\zeta_{{\bf k}\nu}$ are the
eigenvalues and eigenvectors of $\hat{H}_0$.
By using Eqs.~(\ref{relation_1}) and (\ref{relation_2}), one obtains
\begin{eqnarray}
  \nonumber
  &&\hspace{-0.5cm}
  {\rm Re}{\sigma}^{\rm ave}_{xx}(\omega_l)
  =\frac{4\pi}{\omega_l} \hspace{0.cm}
  \sum_{{\bf k} \eta_1\eta_2 \atop N n_1 n_2} \hspace{0cm}
  \langle\phi_{{\bf k}\eta_2}^{n_1-N}| \hat{j}^x_{{\bf k}}
  |\phi_{{\bf k}\eta_1}^{n_1}\rangle  
  \langle\phi_{{\bf k}\eta_1}^{n_2}| \hat{j}^x_{{\bf k}}
  |\phi_{{\bf k}\eta_2}^{n_2-N}\rangle \\ 
  && \hspace{0.9cm}
  \mbox{} \times (f_{{\bf k}\eta_1}^{\rm ini}-f_{{\bf k}\eta_2}^{\rm ini})
  \delta(\varepsilon_{{\bf k}\eta_1}-\varepsilon_{{\bf k}\eta_2}-N\Omega
  +\omega_l)
  \label{opt_ini}
\end{eqnarray} 
with $f_{{\bf k}\eta}^{\rm ini}=\sum\limits_{\nu l}n_{\rm F}(E_{{\bf k}\nu})
  |\langle\zeta_{{\bf k}\nu}|\phi_{{\bf k}\eta}^{l}\rangle|^2$.
It is seen that the above equation is in the same form as 
Eq.~(\ref{opt_ave}), but the distribution 
$f_{{\bf k}\eta}^{\rm ini}$ is quite different from 
$f_{{\bf k}\eta}=n_{\rm F}(\overline{\varepsilon}_{{\bf k}\eta})$ used
in the present paper. 

\section{Approximate analytical  solution of Schr\"odinger
  equation} 
\label{analy_solution}
We first transform the effective Hamiltonian into the basis set formed
by the eigenvectors of $H_0$. Thus the Schr\"odinger equation can be
written as 
\begin{equation}
  \left[v_{\rm F}k \hat{\sigma}_z + \tilde{{H}}_{\rm THz}(t) \right]
  |\Psi_{{\bf k}\eta}(t) \rangle
  =i\frac{\partial}{\partial t}|\Psi_{{\bf k}\eta}(t) \rangle.
  \label{equ_trans}
\end{equation}
Here $|\Psi_{{\bf k}\eta}(t) \rangle =\hat{U}_{{\bf k}}^\dagger
|\Phi_{{\bf k}\eta}(t) \rangle$, with the transformation matrix
\begin{equation}
  U_{\bf k}=\frac{1}{\sqrt{2}}\begin{pmatrix}e^{-i\theta_{\bf k}} &
    -e^{-i\theta_{\bf k}}\\1 & 1\end{pmatrix}; 
  \label{U_k}
\end{equation}
$\tilde{{H}}_{\rm THz}(\theta_{\bf k},t)=\hat{U}_{\bf k}^\dagger
\hat{{H}}_{\rm THz}(t)\hat{U}_{\bf k}$ can be divided into the
intraband $\tilde{{H}}^{\rm intra}_{\rm THz}(\theta_{\bf k},t)$
and interband $\tilde{{H}}^{\rm inter}_{\rm THz}(\theta_{\bf k},t)$ terms, 
given by 
\begin{eqnarray}
  \nonumber
   \tilde{{H}}^{\rm intra}_{\rm THz}(\theta_{\bf k},t)&=&\beta\Omega\hat{\sigma}_z
   (-\cos\theta_{\bf E}\cos\theta_{\bf k}\sin\Omega t \\ &&
   \mbox{}+ \sin\theta_{\bf E}\sin\theta_{\bf k}\cos\Omega t),\\ 
  \label{H_THz_intra}
  \nonumber
  \tilde{{H}}^{\rm inter}_{\rm THz}(\theta_{\bf k},t)&=&\beta\Omega\hat{\sigma}_y
  (\cos\theta_{\bf E}\sin\theta_{\bf k}\sin\Omega t \\ &&
  \mbox{}+ \sin\theta_{\bf E}\cos\theta_{\bf k}\cos\Omega t).
  \label{H_THz_inter}
\end{eqnarray}
Then we solve the Schr\"odinger equation 
without the interband term $\tilde{{H}}^{\rm inter}_{\rm THz}(\theta_{\bf k},t)$
and obtain
\begin{eqnarray}
   |\Psi_{{\bf k}+}^{(0)}(t) \rangle &=& e^{-iv_{\rm F}kt} 
   u^{(0)}(\theta_{\bf k},t) (1,0)^{\rm T}, 
   \label{Psi_0+}\\ 
   |\Psi_{{\bf k}-}^{(0)}(t) \rangle &=& e^{iv_{\rm F}kt} 
   [u^{(0)}(\theta_{\bf k},t)]^\ast (0,1)^{\rm T},
  \label{Psi_0-}
\end{eqnarray}
with 
\begin{eqnarray}
  \nonumber
  u^{(0)}(\theta_{\bf k},t)&=&\exp\left\{i\beta\cos\theta_{\rm E}
    \cos\theta_{\bf k}[1-\cos(\Omega t)]\right.\\
  &&\left. \mbox{} -i \beta\sin\theta_{\rm E}\sin\theta_{\bf k}
    \sin(\Omega t)\right\}.
  \label{u_0}
\end{eqnarray}
It is evident that $u^{(0)}(\theta_{\bf k},t)$ is a time-periodic function,
thus the quasi-energy of $|\Psi_{{\bf k}\eta}^{(0)}(t)\rangle$ 
is exactly the same as the field-free energy $\eta v_{\rm F}k$.
This indicates that all quasi-energy gaps disappear without 
$\tilde{{H}}^{\rm inter}_{\rm THz}(\theta_{\bf k},t)$ . 

The next step is to write the solution of Eq.~(\ref{equ_trans}) into the form
\begin{eqnarray}
  |\Psi_{{\bf k}\eta}(t) \rangle=a_{\eta,1}(t)|\Psi_{{\bf k}+}^{(0)}(t)
  \rangle + a_{\eta,2}(t)|\Psi_{{\bf k}-}^{(0)}(t) \rangle.
  \label{Psi_k_form}
\end{eqnarray}
Substituting Eq.~(\ref{Psi_k_form}) into Eq.~(\ref{equ_trans}), one obtains
\begin{eqnarray}
  && \hspace{-1.1cm}
  \partial_ta_{\eta,1}(t)= -\frac{1}{2} \beta\Omega a_{\eta,2}(t) 
  \sum_l y_l^\ast(\theta_{\bf k}) e^{i(2v_{\rm F}k-l\Omega)t},
  \label{equ_a1}
  \\
  &&\hspace{-1.1cm}
  \partial_ta_{\eta,2}(t)= \frac{1}{2} \beta\Omega a_{\eta,1}(t) 
  \sum_ly_l(\theta_{\bf k}) e^{-i(2v_{\rm F}k-l\Omega)t},
  \label{equ_a2}
\end{eqnarray}
where 
\begin{eqnarray}
  \nonumber
  && \hspace{-1.75cm}
  y_l(\theta_{\bf k})= u_{l-1}^{(1)}(\theta_{\bf k}) 
  ( \sin\theta_{\bf E}\cos\theta_{\bf k} 
  -i\cos\theta_{\bf E}\sin\theta_{\bf k})\\ &&
  \hspace{-0.55cm} \mbox{}+  u_{l+1}^{(1)}(\theta_{\bf k}) 
  (\sin\theta_{\bf E}\cos\theta_{\bf k} 
  + i\cos\theta_{\bf E}\sin\theta_{\bf k}),
  \label{y_n}\\
  \nonumber
  &&\hspace{-1.75cm}
  u_{l}^{(1)}(\theta_{\bf k})=\frac{1}{T_0}\int_0^{T_0}dt 
  \,e^{il\Omega t} [u^{(0)}(\theta_{\bf k},t)]^{2}\\
  &\hspace{0.4cm}=&\hspace{-0.05cm} 
  e^{i2\beta\cos\theta_{\bf E}\cos\theta_{\bf k}}
  J_l(2|z(\theta_{\bf k})|)\left[\frac{-z(\theta_{\bf k})}
  {|z(\theta_{\bf k})|}\right]^l \hspace{-0.1cm},\\
  \label{u_1_l}
  &&\hspace{-1.38cm} 
  z(\theta_{\bf k})=\beta(\sin\theta_{\bf E}\sin\theta_{\bf k}
  +i\cos\theta_{\bf E}\cos\theta_{\bf k}).
  \label{z_0}
\end{eqnarray}
In above derivation, we have applied the summation rule of the Bessel
function\cite{formula}
\begin{equation} 
  J_l(|\mathcal{Z}|)(\frac{\mathcal{Z}^l}{|\mathcal{Z}|^l})
  =\sum_je^{ij\theta}J_j(y)J_{l-j}(x),
  \label{sum_formula}
\end{equation}
with $\mathcal{Z}=x+e^{i\theta}y$ ($x$ and $y$ are real numbers).
Since Eqs.~(\ref{equ_a1}) and (\ref{equ_a2}) cannot be solved in
analytical closed form, 
we solve these equations via the rotating-wave
approximation, which is widely used in the weak electromagnetic field
related problem.\cite{Hanggi_98,Hanggi_05} 
Exploiting this approximation, we neglect the rapidly varying
terms with {$l\ne m$} near the resonant point $2v_{\rm F}k\sim m\Omega$
and obtain
\begin{eqnarray}
  \hspace{-0.5cm}
  \partial_ta_{\eta,1}(t)&=& -\frac{1}{2} \beta\Omega a_{\eta,2}(t)
  y_m^\ast(\theta_{\bf k}) e^{i\delta_{km} t},
  \label{equ_a1_m}
  \\
  \hspace{-0.5cm}
  \partial_ta_{\eta,2}(t)&=& \frac{1}{2} \beta\Omega a_{\eta,1}(t)
  y_m(\theta_{\bf k}) e^{-i\delta_{km} t},
  \label{equ_a2_m}
\end{eqnarray}
with $\delta_{km}=2v_{\rm F}k-m\Omega$.
Comparing the multi-photon resonance equations [Eqs.~(\ref{equ_a1_m})
and (\ref{equ_a2_m})] with the well-known single-photon resonance
equations in the two-level system [e.g., Eq.~(83) in
Ref.~\onlinecite{Hanggi_98}], one can find that the main difference is
the appearance of the coefficient 
$y_m(\theta_{\bf k})$.\cite{little_change} 
Thus $|y_m(\theta_{\bf k})|$ can be treated as a parameter to describe the
magnitude of the effective coupling induced by the $m$-photon resonances. 

\begin{figure}[tbp]
  \begin{center}
    \includegraphics[width=6.2cm]{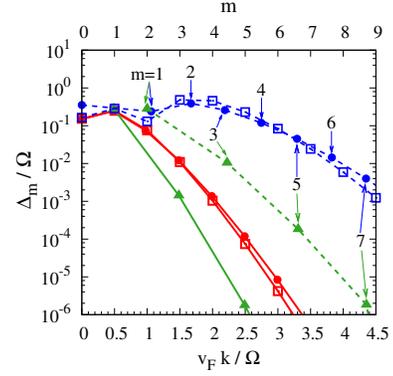}
  \end{center}
  \caption{ (Color online) Quasi-energy gap $\Delta_m$ versus the
    corresponding momentum 
    from the exact calculation (dots) and the approximate formula 
    Eq.~(\ref{gap_approx}) (squares) under circularly polarized THz
    fields with $\beta=0.4$ (solid curves) and $2.3$ (dashed curves). 
    The triangles represent the results from the exact calculation
    without the intraband term $\tilde{H}^{\rm intra}_{\rm THz}$.
    The indices  $m$ of the gaps at finite momentum from the exact
    calculation 
    for $\beta=2.3$ are labelled in the figure. The other gap indices
    can be read from the $m$-axis on the upper frame. 
    Note that gaps only appear at the momentums where the dots
    (squares, triangles) indicate. The curves are only plotted as a
    guide for the eyes.
  }
  \label{fig_gap}
\end{figure}

The solutions of Eqs.~(\ref{equ_a1_m}) and (\ref{equ_a2_m}) are then
easy to obtain: 
\begin{eqnarray}
  &&\hspace{-1.2cm}
  a_{\eta,1}=C_\eta^+ \exp\left\{\frac{it}{2}\left[\delta_{km}-\eta\sqrt{\delta_{km}^2
        +[\Delta^\prime_m(\theta_{\bf k})]^2}\right]\right\},
  \label{a1_final}
  \\ && \hspace{-1.2cm}
  a_{\eta,2}=C_\eta^- \exp\left\{-\frac{it}{2}\left[\delta_{km}+\eta\sqrt{\delta_{km}^2
        +[\Delta^\prime_m(\theta_{\bf k})]^2}\right]\right\}.
  \label{a2_final}
\end{eqnarray} 
Here 
\begin{equation}
  \Delta^\prime_m(\theta_{\bf k})=\beta|y_m(\theta_{\bf k})|\Omega;
\end{equation}
$C_\eta^+$ and $C_\eta^-$ are the normalization coefficients satisfying
\begin{eqnarray}
  &&|C_\eta^+|^2+|C_\eta^-|^2=1,\\
  &&\frac{C_\eta^+}{C_\eta^-}=-i\frac{\delta_{km} 
    + \eta \sqrt{\delta_{km}^2 + [\Delta^\prime_m(\theta_{\bf k})]^2}}
  {\beta\Omega y_m(\theta_{\bf k})}.
\end{eqnarray} 
Thus one has
\begin{eqnarray}
  \nonumber
   |\Psi_{{\bf k}\eta}(t) \rangle &=& e^{-\frac{i}{2}\eta t\sqrt{\delta_{km}^2
       + [\Delta^\prime_m(\theta_{\bf k})]^2}} \Big( C_\eta^+ e^{-\frac{i}{2}m\Omega t}
   u^{(0)}(\theta_{\bf k},t),
   \\ &&
   C_\eta^-e^{\frac{i}{2}m\Omega t}[u^{(0)}(\theta_{\bf k},t)]^\ast \Big)^{\rm T}.
   \label{Psi_k}
\end{eqnarray} 
Evidently, the corresponding quasi-energy is
\begin{equation}
  \varepsilon_{{\bf k}\eta}=\frac{\eta}{2}\sqrt{\delta_{km}^2
    +[\Delta^\prime_m(\theta_{\bf k})]^2}.
  \label{energy_app}
\end{equation}
Thus the quasi-energy gap at the resonant point 
$\delta_{km}=0$ reads
\begin{equation}
  \Delta_m(\theta_{\bf k})=\beta|y_m(\theta_{\bf k})|\Omega
  \pmod{\Omega}.
  \label{gap_approx}
\end{equation}
It is seen that the magnitude of the gap 
$\Delta_m(\theta_{\bf k})$ is determined
 by the effective coupling parameter $|y_m(\theta_{\bf k})|$.

In Fig.~\ref{fig_gap}, we plot the magnitude of the quasi-energy gaps
$\Delta_m$ against the corresponding momentums from
the exact calculation (dots) and the approximate formula 
Eq.~(\ref{gap_approx}) (squares) under circularly polarized THz
fields with different field strengths. 
As shown in Figs.~\ref{fig_energy_strong}(a) and (b), the
momentums of the gaps from the calculation markedly deviate from 
$m\Omega/(2v_{\rm F})$ in the strong field regime. Therefore, 
we define $\Delta_0$ as the gap at $k=0$ and $\Delta_m$ ($m\ne0$) as
the $m$-th gap from the one at $k=0$ for gaps from the exact
calculation.  The corresponding indices $m$ are labelled in Fig.~\ref{fig_gap}.
In the case with low field strength (red solid curves in
Fig.~\ref{fig_gap}), it is seen that the results from the
approximate formula agree well with the ones from the exact computation.
In the case with high field strength (blue dashed curves), the
magnitude of the gaps from these two approaches are comparable,
but the corresponding momentums differ significantly,
owing to the rotating-wave approximation.
In Fig.~\ref{fig_gap}, we also plot the results from the exact
calculation without the intraband term 
$\tilde{H}^{\rm intra}_{\rm THz}$
(triangles). It is seen that except for the gap with $m=1$,
the gaps without $\tilde{H}^{\rm intra}_{\rm THz}$ are much smaller
than the ones with the intraband term.
In particular, the gaps with $m$ being even numbers
vanish within the error range of our computation (about
$10^{-10}\Omega$). These results indicate that the intraband term 
$\tilde{H}^{\rm intra}_{\rm THz}$ plays a significant role in the
formation of the quasi-energy gaps induced by the multi-photon
resonances. 

\begin{figure}[tbp]
  \begin{center}
    \includegraphics[width=6.2cm]{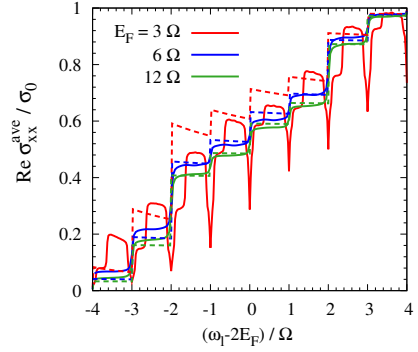}
  \end{center}
  \caption{ (Color online) Time-averaged optical conductivities from
    the exact calculation (solid curves) and the the approximate
    formula Eq.~(\ref{ana_fin_dis2}) (dashed curves) are plotted as
    function of the optical frequency under a circularly polarized THz
    field with $\beta=2.3$ for different Fermi energies. 
  }
  \label{fig_cir_hEf2}
\end{figure}

\section{Approximate analytical formula of optical conductivity}
\label{analy_opt}
Now we discuss the approximate analytical formula of the optical
conductivity at high Fermi energy.
It is known that the optical conductivity in this case is only from
the contribution from the states with large momentum, where the
effect from the interband term of ${H}_{\rm THz}$ becomes
negligible. Thus one can neglect the interband term and
obtain 
the eigenvector
\begin{equation}
  |\Phi_{{\bf k}\eta}(t)\rangle=\hat{U}_{{\bf k}}
  |\Psi_{{\bf k}\eta}^{(0)}(t)\rangle
\end{equation}
in which $U_{\bf k}$ is given by Eq.~(\ref{U_k}) and
$|\Psi_{{\bf k}\eta}^{(0)}(t)\rangle$ is determined by
Eqs.~(\ref{Psi_0+}) and (\ref{Psi_0-}). Then one has
\begin{eqnarray}
  \overline{\varepsilon}_{{\bf k}\eta}&=&\varepsilon_{{\bf k}\eta} 
  =\eta v_{\rm F}k,
  \label{E_modified}\\
  |\phi_{{\bf k}+}^{n}\rangle &=& u^{(0)}_{n}(\theta_{\bf k})
  (e^{-i\theta_{\bf k}},1)^{\rm T}/\sqrt{2}, \\ 
  |\phi_{{\bf k}-}^{n}\rangle &=& [u^{(0)}_{-n}(\theta_{\bf k})]^\ast
  (-e^{-i\theta_{\bf k}},1)^{\rm T}/\sqrt{2},
  \label{phi_modified}
\end{eqnarray}
where 
\begin{eqnarray}
  \nonumber
   &&\hspace{-0.8cm} u^{(0)}_n(\theta_{\bf k}) =
    e^{i\beta\cos\theta_{\bf E}\cos\theta_{\bf k}}
  J_n(|z(\theta_{\bf k})|)\left[\frac{-z(\theta_{\bf k})}
    {|z(\theta_{\bf k})|}\right]^n
  \label{u_0_m}
\end{eqnarray}
with $z(\theta_{\bf k})$ given by Eq.~(\ref{z_0}).
Substituting Eqs.~(\ref{E_modified})-(\ref{u_0_m}) into
Eq.~(\ref{opt_ave}), one obtains
\begin{eqnarray}
  \nonumber && \hspace{-0.4cm}
  {\rm Re}{\sigma}^{\rm ave}_{ll}(\omega_l) =
  \frac{e^2}{4\pi} \sum_{N} \hspace{-0.cm}
  (1-N\frac{\Omega}{\omega_l}) R_N \theta({\omega_l}-2E_{\rm F}-N{\Omega}), \\ &&
  \label{ana_fin_dis2}
\end{eqnarray}
where 
\begin{eqnarray}
  \nonumber
  R_N&=& \int_0^{2\pi} d\theta_{\bf k} \mathcal{A}_l^2(\theta_{\bf k})
  J_N^2(2|z(\theta_{\bf k})|),
  \label{R_m}
\end{eqnarray}
with $\mathcal{A}_x(\theta_{\bf k})=\sin\theta_{\bf k}$ and 
$\mathcal{A}_y(\theta_{\bf k})=\cos\theta_{\bf k}$. 

In Fig.~\ref{fig_cir_hEf2}, we plot the optical conductivities
from the exact calculation (solid curves) and the the approximate formula
Eq.~(\ref{ana_fin_dis2}) (dashed curves) as function of the frequency
for different Fermi energies. 
It is seen that the difference between two calculations becomes
smaller when the Fermi energy increases. This indicates that the
effect from the interband term of ${H}_{\rm THz}$ becomes negligible
for high Fermi energy, in consistence with the discussion in the main
text.

\end{appendix}


\begin{thebibliography}{0}
\bibitem{Novoselov_04} K. S. Novoselov, A. K. Geim, S. V. Morozov, D. Jiang,
  Y. Zhang, S. V. Dubonos, I. V. Grigorieva, and A. A. Firsov, Science
  {\bf 306}, 666 (2004).
\bibitem{Geim_rev_nat} A. K. Geim and K. S. Novoselov, Nature Mater. {\bf 6}, 183
  (2007).
\bibitem{Neto_rev_09} A. H. Castro Neto, F. Guinea, N. M. R. Peres,
  K. S. Novoselov, and A. K. Geim, Rev. Mod. Phys. {\bf 81}, 109 (2009).
\bibitem{Beenakker_rev} C. W. J. Beenakker, Rev. Mod. Phys. {\bf 80}, 1337 (2008).
\bibitem{Chakraborty_rev} D. S. L. Abergel, V. Apalkov, J. Berashevich,
  K. Ziegler, and T. Chakraborty, Adv. Phys. {\bf 59}, 261 (2010). 
\bibitem{Peres_rev} N. M. R. Peres, Rev. Mod. Phys. {\bf 82}, 2673 (2010).
\bibitem{Mucciolo_rev} E. R. Mucciolo and C. H. Lewenkopf, J. Phys.:
  Condens. Matter {\bf 22}, 273201 (2010).
\bibitem{Sarma_rev} S. Das Sarma, S. Adam, E. H. Hwang, and E. Rossi,
  Rev. Mod. Phys., 2011, in press (also arXiv:1003.4731).
\bibitem{Ferreira_dc} A. Ferreira, J. Viana-Gomes, J. Nilsson, E. R. Mucciolo,
  N. M. R. Peres, and A. H. Castro Neto, arXiv:1010.4026.
\bibitem{Orlita_rev}M. Orlita and M. Potemski, Semicond. Sci. Technol. {\bf 25},
  063001 (2010).
\bibitem{Stauber_NN} T. Stauber, N. M. R. Peres, and A. K. Geim, Phys. Rev. B
  {\bf 78}, 085432 (2008).
\bibitem{Ando_02} T. Ando, Y. Zheng, and H. Suzuura, J. Phys. Soc. Jpn. {\bf
    71}, 1318 (2002).
\bibitem{Gusynin} V. P. Gusynin, S. G. Sharapov, and J. P. Carbotte,
  Phys. Rev. Lett. {\bf 96}, 256802 (2006).
\bibitem{Neto_06} N. M. R. Peres, F. Guinea, and A. H. Castro Neto, Phys. Rev. B
  {\bf 73}, 125411 (2006).
\bibitem{Stauber_disorder} T. Stauber, N. M. R. Peres, and A. H. Castro Neto,
  Phys. Rev. B {\bf 78}, 085418 (2008).
\bibitem{Giuliani_ee} A. Giuliani, V. Mastropietro, and M. Porta,
  arXiv:1101.2169.
\bibitem{exp_Basov} Z. Q. Li, E. A. Henriksen, Z. Jiang, Z. Hao, M. C. Martin,
  P. Kim, H. L. Stormer, and D. N. Basov, Nat. Phys. {\bf 4}, 532 (2008).
\bibitem{exp_Nair} R. R. Nair, P. Blake, A. N. Grigorenko, K. S. Novoselov,
  T. J. Booth, T. Stauber, N. M. R. Peres, and A. K. Geim, Science {\bf 320},
  1308 (2008).
\bibitem{exp_APL} J. M. Dawlaty, S. Shivaraman, J. Strait, P. George,
  M. Chandrashekhar, F. Rana, M. G. Spencer, D. Veksler, and Y. Chen,
  Appl. Phys. Lett. {\bf 93}, 131905 (2008).
\bibitem{Syzranov_08} S. V. Syzranov, M. V. Fistul, and K. B. Efetov, Phys. Rev. B
  {\bf 78}, 045407 (2008).
\bibitem{Oka_cur} T. Oka and H. Aoki, Phys. Rev. B {\bf 79}, 081406(R) (2009);
  {\it ibid.} {\bf 79}, 169901(E) (2009).
\bibitem{Oka_cur2} T. Oka and H. Aoki, J. Phys.: Conf. Ser. {\bf 200}, 062017
  (2010). 
\bibitem{Zhang_energy} W. Zhang, P. Zhang, S. Duan, and X. Zhao, New
  J. Phys. {\bf 11}, 063032 (2009). 
\bibitem{Kibis_quantiz} O. V. Kibis, Phys. Rev. B {\bf 81}, 165433 (2010).
\bibitem{Naumis_pol} F. J. L\'{o}pez-Rodr\'{i}guez and G. G. Naumis,
  Phys. Rev. B {\bf 78}, 201406(R) (2008).
\bibitem{Oka_opt} T. Oka and H. Aoki, arXiv:1007.5399.
\bibitem{Fistul_07} M. V. Fistul and K. B. Efetov, Phys. Rev. Lett. {\bf 98},
  256803 (2007).
\bibitem{Gupta_dis2} A. K. Gupta, O. E. Alon, and N. Moiseyev, Phys. Rev. B
  {\bf 68}, 205101 (2003).
\bibitem{Mikhailov_multi} S. A. Mikhailov and K. Ziegler,
  J. Phys.: Condens. Matter {\bf 20}, 384204 (2008).
\bibitem{Wright_09} A. R. Wright, X. G. Xu, J. C. Cao, and C. Zhang,
  Appl. Phys. Lett. {\bf 95}, 072101 (2009).
\bibitem{Ryzhii_dc} F. T. Vasko and V. Ryzhii, Phys. Rev. B {\bf 77}, 195433 (2008).
\bibitem{Ryzhii_dcac} A. Satou, F. T. Vasko, and V. Ryzhii, Phys. Rev. B {\bf 78},
  115431 (2008).
\bibitem{Ryzhii_BG} V. Ryzhii and M. Ryzhii, Phys. Rev. B {\bf 79}, 245311 (2009).
\bibitem{Wright_BG} A. R. Wright, J. C. Cao, and C. Zhang, Phys. Rev. Lett.
  {\bf 103}, 207401 (2009).
\bibitem{Abergel_BG} D. S. L. Abergel and T. Chakraborty, Appl. Phys. Lett.
  {\bf 95}, 062107 (2009).
\bibitem{Abergel_BG2} D. S. L. Abergel and T. Chakraborty,
  Nanotechnology {\bf 22}, 015203 (2011).
\bibitem{Kono_sideband} J. \v{C}erne, J. Kono, T. Inoshita, M. Sherwin, M. Sundaram, and
  A. C. Gossard, Appl. Phys. Lett. {\bf 70}, 3543 (1997); J. Kono, M. Y. Su,
  T. Inoshita, T. Noda, M. S. Sherwin, S. J. Allen, Jr., and H. Sakaki,
  Phys. Rev. Lett. {\bf 79}, 1758 (1997).
\bibitem{Nordstrom_exp_EDFK} K. B. Nordstrom, K. Johnsen, S. J. Allen,
  A.-P. Jauho, B. Birnir, J. Kono, T. Noda, H. Akiyama, and H. Sakaki,
  Phys. Rev. Lett. {\bf 81}, 457 (1998).
\bibitem{Phillips_sideband} C. Phillips, M. Y. Su, M. S. Sherwin, J. Ko, and
  L. Coldren, Appl. Phys. Lett. {\bf 75}, 2728 (1999).
\bibitem{Maslov_sideband} A. V. Maslov and D. S. Citrin, Phys. Rev. B {\bf 62},
  16686 (2000).
\bibitem{Holthaus_Stark} M. Holthaus and D. W. Hone, Phys. Rev. B {\bf 49},
  16605 (1994).
\bibitem{Rodriguez_Stark} A. H. Rodr\'{\i}guez, L. Meza-Montes, C. Trallero-Giner,
  and S. E. Ulloa, Phys. Stat. Sol. (b) {\bf 242}, 1820 (2005). 
\bibitem{Yacoby_68} Y. Yacoby, Phys. Rev. {\bf 169}, 610 (1968).
\bibitem{Jauho_DFK} A.-P. Jauho and K. Johnsen, Phys. Rev. Lett. {\bf 76}, 4576
  (1996); K. Johnsen and A.-P. Jauho, Phys. Rev. B {\bf 57}, 8860 (1998).
\bibitem{Chin_exp_20} A. H. Chin, J. M. Bakker, and J. Kono, Phys. Rev. Lett.
  {\bf 85}, 3293 (2000); A. H. Chin, O. G. Calder\'on, and J. Kono,
  Phys. Rev. Lett. {\bf 86}, 3292 (2001).
\bibitem{Srivastava_exp_DFK}  A. Srivastava, R. Srivastava, J. Wang, and
  J. Kono, Phys. Rev. Lett. {\bf 93}, 157401 (2004).
\bibitem{Zhang_exp_06} T. Y. Zhang and W. Zhao, Phys. Rev. B {\bf 73}, 245337
  (2006).
\bibitem{Hanggi_98} M. Grifoni and P. H\"anggi, Phys. Rep. {\bf 304}, 229
  (1998).
\bibitem{Hanggi_05} S. Kohler, J. Lehmann, and P. H\"anggi, Phys. Rep.
  {\bf 406}, 379 (2005).
\bibitem{Cheng_APL} J. L. Cheng and M. W. Wu, Appl. Phys. Lett.
  {\bf 86}, 032107 (2005).
\bibitem{Jiang_JAP} J. H. Jiang, M. Q. Weng, and M. W. Wu, J. Appl. Phys.
  {\bf 100}, 063709 (2006).
\bibitem{Zhou_PE} Y. Zhou, Physica E {\bf 40}, 2847 (2008).
\bibitem{Jiang_PRB_07} J. H. Jiang and M. W. Wu, Phys. Rev. B {\bf 75}, 035307
  (2007).
\bibitem{Jiang_PRB_08} J. H. Jiang, M. W. Wu, and Y. Zhou, Phys. Rev. B
  {\bf 78}, 125309 (2008).
\bibitem{Shirley_65}  J. H. Shirley, Phys. Rev. {\bf 138}, B979 (1965).
\bibitem{Haug_08} H. Haug and A.-P. Jauho, {\it Quantum kinetics in
    Transport and Optics of Semiconductors} (Springer, Berlin, 2008).
\bibitem{DiVincenzo} D. P. DiVincenzo and E. J. Mele, Phys. Rev. B {\bf 29},
  1685 (1984).    
\bibitem{other_def} $\beta$ in this paper is equivalent to the dimensionless
  quantity $\sqrt{2}F/\Omega$ in the previous investigations,\cite{Oka_cur,Oka_cur2,
    Oka_opt} in which $F=eE_0a/\sqrt{2}$ and $\Omega=v_{\rm F}/a$
  with $a$ being the lattice constant.
\bibitem{Faisal_dis2} F. H. M. Faisal and J. Z. Kami\'nski, Phys. Rev. A
  {\bf 54}, R1769 (1996); {\it ibid.} {\bf 56}, 748 (1997).
\bibitem{Hsu_dis2} H. Hsu and L. E. Reichl, Phys. Rev. B {\bf 74},
  115406 (2006).  
\bibitem{Martinez_meanenergy} D. F. Martinez, L. E. Reichl, and
  G. A. Luna-Acosta, Phys. Rev. B {\bf 66}, 174306 (2002).
\bibitem{Kohler_PRE} S. Kohler, T. Dittrich, and P. H\"anggi, Phys. Rev. E
  {\bf 55}, 300 (1997).
\bibitem{Wu_rev} M. W. Wu, J. H. Jiang, and M. Q. Weng,
  Phys. Rep. {\bf 493}, 61 (2010), and references therein.
\bibitem{gamma} For the sake of the convergence, we replace the delta
  function $\delta(x)$ in the DOS and optical conductivity by a
  Lorentzian $\frac{\Gamma}{\pi(x^2+\Gamma^2)}$. The values of
  $\Gamma$ are set as $0.001\Omega$ and $0.02\Omega$ for the
  calculation of the DOS and optical conductivity, respectively. It is
  noted that our main results are independent of the value of $\Gamma$
  when $\Gamma$ is small enough. 
\bibitem{gap_zerok} Note that the so-called $0$-photon resonance,
  which leads to the gap at zero momentum, is corresponding to the
  multi-photon process which first absorbs a photon and then emits 
  one or vice versa. 
\bibitem{sing_zerok} Since 
  $\lim\limits_{k\to 0}k/|\nabla_{\bf k}\varepsilon_{{\bf k}\eta}|$  
  is finite, the van Hoff singularity at $k=0$ does not induce any
  divergence in the DOS. However, the states around zero momentum
  can still significantly enhance the DOS at the frequencies of the
  sidebands with large weight, as shown in the DOS limited in
  $k<0.5\Omega$ [green dotted curve in Fig.~\ref{fig_DOS}(b)]. 
\bibitem{van_Hoff} The DOS corresponding to the van Hoff singularities 
  diverges as ${\omega}^{-1/2}$ in the case with isotropic energy
  spectrum, and diverges logarithmically in the case with anisotropic
  energy spectrum. See also [P. Y. Yu and M. Cardona, {\em
    Fundamentals of Semiconductors} (Springer, Berlin, 2005)].
\bibitem{angle_symmetry} It is easy to prove that the quasi-energies
  and mean energies for $\pm \theta_{\bf k}$ and 
  $\pi \pm \theta_{\bf k}$ are identical.
\bibitem{little_change} Comparing the Hamiltonian of the external
  field in this paper with the one in Ref.~\onlinecite{Hanggi_98},
  one can see $\beta\Omega$  corresponds to $\epsilon/2$ in
  Ref.~\onlinecite{Hanggi_98}. 
\bibitem{formula} I. S. Gradshteyn and I. M. Ryzhik, {\em Table of
  Integrals, Series and Products}, 5th ed. (Academic, New York, 1994).
\end{thebibliography}
\end{document}